\pdfoutput=1
\documentclass[11pt]{article}

\usepackage{PRIMEarxiv}
 \usepackage{wrapfig}
\usepackage[utf8]{inputenc}
\usepackage[T1]{fontenc}

\usepackage[expansion=false]{microtype}
\usepackage{amsmath,amssymb,amsthm}
\usepackage{mathtools}
\usepackage{graphicx}
\usepackage{booktabs}
\usepackage{multirow}
\usepackage{array}
\usepackage{xcolor}
\usepackage{enumitem}
\usepackage{caption}
\usepackage[caption=false]{subfig}
\usepackage{algorithm}
\usepackage{algpseudocode}
\usepackage{adjustbox}   % shrinks a table only if it is wider than the text
\usepackage{xurl}
\usepackage[numbers,sort&compress]{natbib}
\usepackage[colorlinks=true,linkcolor=blue!50!black,citecolor=green!40!black,urlcolor=blue!50!black]{hyperref}
\usepackage[capitalise,nameinlink]{cleveref}

\graphicspath{{figures/}}

\theoremstyle{plain}
\newtheorem{theorem}{Theorem}[section]
\newtheorem{proposition}[theorem]{Proposition}
\newtheorem{lemma}[theorem]{Lemma}
\newtheorem{corollary}[theorem]{Corollary}
\theoremstyle{definition}
\newtheorem{definition}[theorem]{Definition}

\theoremstyle{remark}
\newtheorem{remark}[theorem]{Remark}

\newcommand{\D}{\mathcal{D}}

\newcommand{\sinf}{\sigma_\infty}
\newcommand{\bone}{\beta_1}
\newcommand{\Hg}{\mathcal{H}}

\newcommand{\nloc}{\lambda}
\newcommand{\hfrac}{h}

\title{{Dessins d'Enfants as a Topological Lens on Temporal Hypergraphs}}
\author{
  Mehmet Emin Aktas \\
  School of Data Science and Analytics \\
  Kennesaw State University \\
  Kennesaw, GA, USA \\
  \texttt{maktas1@kennesaw.edu} \\
  \And
  Esra Akbas \\
  Department of Computer Science \\
  Georgia State University \\
  Atlanta, GA, USA \\
  \texttt{eakbas1@gsu.edu} \\
}
\date{}

\begin{document}

\maketitle

\begin{abstract}
Hypergraphs are the natural language for group interactions, and topological data analysis has given us several ways to look at them: simplicial and cell complexes, persistent homology, and sheaves. All of these treat a hyperedge as an unordered set. In this paper we introduce a different and, to our knowledge, new viewpoint that comes from Grothendieck's theory of dessins d'enfants and, unlike these approaches, takes into account the order in which vertices appear in each hyperedge and hyperedges appear around each vertex. A dessin is a bipartite graph drawn on a closed orientable surface. Through Walsh's classical correspondence between hypermaps and bipartite maps, every hypergraph equipped with a cyclic order of incidences at each vertex and inside each hyperedge \emph{is} a dessin, and therefore determines a surface with faces and a genus. We develop this idea for temporal hypergraphs and build the surface by replaying history: incidences are added in the order in which they occurred, and each new incidence is inserted into the cyclic orders built so far, next to the earlier incidence to which it is related. We prove that every insertion either grows the hypergraph without creating a cycle, closes a cycle \emph{locally} by splitting a face, or closes it \emph{globally} by adding a handle; the number of independent cycles equals the number of local plus handle events, and the genus equals the number of handle events. These counts give interpretable, size-controlled measures of how a temporal hypergraph's cycles organize on a surface as it grows. We apply our method to five Stack Exchange communities and show that it separates communities that standard statistics cannot tell apart, and quantifies how much the real order of interactions matters. We further discuss what the surface viewpoint adds to existing higher-order network analysis and how it opens a path toward dessin-based learning on hypergraphs.
\end{abstract}


%=====================================================================================================
\section{Introduction}
\label{sec:intro}

Many systems are made of interactions among more than two parties at once: co-authored papers, e-mail threads, chemical reactions, meetings, and online discussions. Hypergraphs, in which an edge may contain any number of vertices, are the standard model for such data~\citep{Berge1973,Battiston2020,Bick2023,Torres2021,kusakula2025beyond}. Over the last decade, ideas from algebraic topology have become an important part of the toolbox for analyzing them. A hypergraph can be completed into a simplicial complex and studied with homology and persistent homology~\citep{Aktas2019,aktas2026persistent,gasparovic2026survey}; it can be lifted to a cell or combinatorial complex for topological deep learning~\citep{Hajij2022}; and data can be attached to its incidences with cellular sheaves and their Laplacians~\citep{HansenGhrist2019,Bodnar2022}. Hypergraph neural networks~\citep{Feng2019} and incidence-based models with edge-dependent weights~\citep{ChitraRaphael2019,Wang2023EDHNN} complete the picture on the learning side.

A common feature of these approaches is that a hyperedge is an \emph{unordered set}. Yet in many applications, the incidences carry a natural order: the answers in a question--answer thread arrive one after another and reply to specific earlier posts, the authors of a paper are listed in order, and the hyperedges around a vertex are ordered in time, as when a forum user joins one thread after another. This paper asks what topology such orders carry, and answers it with a classical but, in data science, unused mathematical object: the \emph{dessin d'enfant}.

Informally, a dessin d'enfant (French for ``child's drawing'') is a graph drawn on a surface, with vertices colored black and white so that every edge joins a black vertex to a white one. Despite this simple appearance, dessins are central objects in modern mathematics. Grothendieck introduced them in the 1980s after Belyi's theorem showed that such drawings encode algebraic curves, which connects them to number theory and Galois theory~\citep{Belyi1979,Grothendieck1997Esquisse}. For our purposes, only their combinatorial description matters: a dessin is the same as a pair of permutations $(\sigma_0,\sigma_1)$ on a finite set of edges, called \emph{darts}. The permutation $\sigma_0$ records the cyclic order of the darts around each black vertex, $\sigma_1$ the cyclic order around each white vertex, and their product $\sinf=(\sigma_0\sigma_1)^{-1}$ walks around the faces of the drawing~\citep{LandoZvonkin2004,JonesWolfart2016,GirondoGonzalezDiez2012,Schneps1994}. Walsh observed that the same data describe a \emph{hypermap}, that is, a hypergraph drawn on a surface: black vertices are the vertices of the hypergraph, white vertices are its hyperedges, and darts are incidences~\citep{Walsh1975,Cori1975,JonesSingerman1978}.

The correspondence above means that a hypergraph becomes a surface as soon as we choose, for every vertex and every hyperedge, a cyclic order of its incidences. Once the orders are fixed, everything else is forced: the faces, the number of connected components, the Euler characteristic and the genus. The surface records how the cycles of the hypergraph fit together. We show (\cref{prop:cycles}) that the cycle rank $\bone$ of the hypergraph splits exactly into \emph{local} cycles, which bound faces, and \emph{global} cycles, which run around handles \[ \bone \;=\; (F-c) \;+\; 2g, \] where $F$ is the number of faces, $c$ the number of connected components and $g$ the genus. The split depends on the cyclic orders, and that is where the information lies.

For temporal hypergraphs, the obvious choice is to read the cyclic orders directly off the timestamps: the hyperedges of a vertex in the order the vertex joined them, and the vertices of a hyperedge in the order they joined it. This fails. We show in \cref{sec:negative} that such rotations behave exactly like random rotations: the surfaces they produce close essentially no cycles locally. The reason is simple. A face closes only if consecutive incidences line up at \emph{every} vertex and hyperedge it passes, and the order of incidences inside a hyperedge and the order of hyperedges around a vertex are produced by unrelated processes, each read off independently of the other.

We therefore introduce the \emph{growing dessin} (\cref{sec:online}), which applies to any temporal hypergraph and uses time differently: time determines the \emph{order in which incidences are inserted}, not the cyclic orders themselves. Each incidence inserts one new dart into the existing cyclic orders, which are never rearranged, and \emph{where} it goes is decided by the surface built so far and by its \emph{anchor}, the earlier incidence to which it is related (for example, the post it replies to). The resulting cyclic orders are generally not time orders. We prove that every insertion falls into exactly one of five types with a precisely known effect on the surface (\cref{thm:events}), which gives a dynamic, interpretable topological description of a growing hypergraph.

We apply the growing dessin method to general temporal hypergraphs in Stack Exchange Q\&A communities as our first application domain. We represent users as vertices and threads as hyperedges. We use the posts' timestamps and the reply structure in defining the dessin for each community and show that different communities have very different interaction cultures.

Our contributions are fourfold. We introduce dessins d'enfants (hypermaps) as a representation of hypergraphs with ordered incidences and state the basic theory in the language of data analysis, including the cycle decomposition $\bone=(F-c)+2g$ and its homological meaning (\cref{sec:background,sec:static}). We define the growing dessin for any temporal hypergraph, prove an exact event classification with a conservation law (\cref{thm:events,cor:conservation}), and give an implementation that never rescans large faces (\cref{sec:algorithm}). We define interpretable, size-controlled measures (\cref{sec:measures}) and show that cyclic orders read directly from the data are topologically indistinguishable from random ones (\cref{sec:negative}). Finally, we apply the method to five Stack Exchange communities, including two that standard statistics cannot tell apart, but the dessin can, and test the conclusions against the insertion rule, the temporal order, and an offline ceiling (\cref{sec:results}).

\paragraph{Organization.} \Cref{sec:background} reviews hypergraphs, dessins and hypermaps. \Cref{sec:static} develops the static theory of hypergraph dessins with a small example and relates the dessin to the simplicial completion of a hypergraph. \Cref{sec:online} introduces the growing dessin, proves the event classification, presents the algorithm, defines the measures, discusses other application domains, and works through a toy example step by step. \Cref{sec:setup} describes the data and the experimental protocol. \Cref{sec:results} reports the results, including a negative control, robustness checks, an ablation, and a bracketing of the results between a floor and a ceiling. \Cref{sec:discussion} discusses what dessins add, the limitations, and future directions, and \cref{sec:conclusion} concludes. \Cref{app:impl} gives implementation details.

%=====================================================================================================
\section{Background}
\label{sec:background}

\subsection{Hypergraphs and temporal hypergraphs}
A \emph{hypergraph} $\Hg=(U,T,\D)$ consists of a set $U$ of vertices, a set $T$ of hyperedges and an incidence
relation $\D\subseteq U\times T$; an incidence $(u,t)\in\D$ means that vertex $u$ belongs to hyperedge $t$. Depending on the domain, vertices may be people, molecules or variables, and hyperedges may be discussions,
papers, e-mails, meetings or reactions. To keep the exposition concrete, we often use the vocabulary of our
application, discussion forums, where vertices are \emph{users}, hyperedges are \emph{threads}, and an incidence means that a user participated in a thread; all definitions and results hold for arbitrary hypergraphs. The \emph{incidence graph} $B(\Hg)$ is the bipartite graph with vertex set $U\sqcup T$ and one edge for every incidence $(u,t)\in\D$. We call these edges \emph{darts}. Its cycle rank, or
first Betti number, is
\begin{equation}
  \bone(\Hg) \;=\; |\D| - |U| - |T| + c,
  \label{eq:beta1}
\end{equation}
where $c$ is the number of connected components. $\bone$ counts the independent cycles of the hypergraph; for
example, two vertices that belong together to two different hyperedges create one cycle
$u_1\to t_1\to u_2\to t_2\to u_1$ (two users who meet in two threads, two authors who write two papers together).

A \emph{temporal hypergraph} additionally records when each incidence appeared, so its incidences form a sequence $(u_1,t_1),(u_2,t_2),\dots$ ordered by time. Many temporal hypergraphs also record \emph{relations} between incidences: a forum post replies to an earlier post, an e-mail replies to an earlier message, a new participant is invited to a meeting by an earlier participant. When available, we use such a relation to designate, for each new incidence, an earlier incidence of the same hyperedge that it is attached to, its \emph{anchor} (\cref{sec:online}). In our application, the incidence of a user in a thread is recorded at the time of the user's \emph{first} post in the thread, and its anchor is the incidence of the author of the post it replies to
(\cref{sec:setup}).

\subsection{Dessins d'enfants}
\begin{definition}[Dessin, combinatorial form]
A \emph{dessin} (or \emph{hypermap}) on a finite set $\D$ of darts is a pair of permutations
$(\sigma_0,\sigma_1)$ of $\D$. Its \emph{black vertices} are the cycles of $\sigma_0$, its \emph{white
vertices} the cycles of $\sigma_1$, and its \emph{faces} the cycles of $\sinf \;=\; (\sigma_0\sigma_1)^{-1}, \text{so that } \sigma_0\,\sigma_1\,\sinf=\mathrm{id}.$ We compose permutations from right to left: $(\sigma_0\sigma_1)(d)=\sigma_0(\sigma_1(d))$. The
\emph{components} of the dessin are the orbits of the group $\langle\sigma_0,\sigma_1\rangle$, called the
\emph{monodromy} or \emph{cartographic group}. A dessin is \emph{connected} if this group acts transitively.
\end{definition}

\begin{wrapfigure}{r}{0.34\textwidth}
\vspace{-12pt}
\centering
\includegraphics[width=0.32\textwidth]{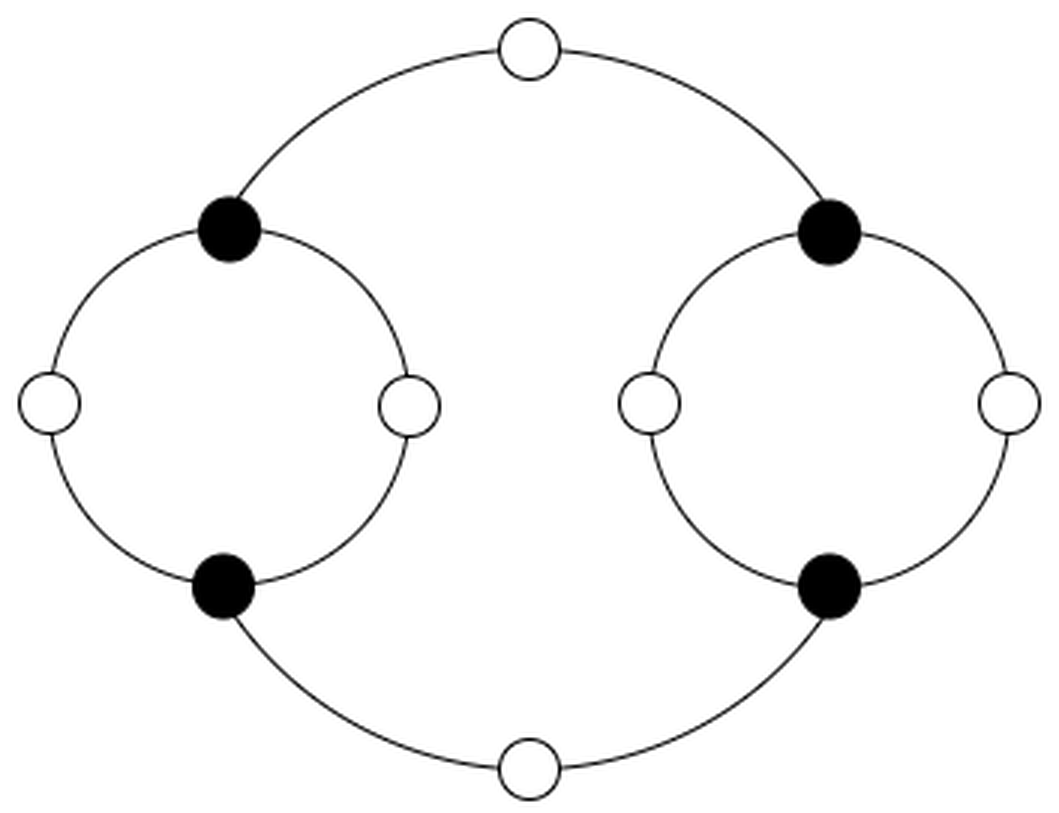}
\caption{An example of a dessin d'enfant, reproduced from~\citep{aktas2017dessins}. It lies on the sphere and has 4 black vertices (each of degree 3), 6 white vertices (each of degree 2), 12 darts and 4 faces: the inside of each circle, the region between the circles, and the outer region. }
\label{fig:dessin-example}
\vspace{-10pt}
\end{wrapfigure}
Geometrically, a dart is an edge of a bipartite graph drawn on an oriented surface; $\sigma_0(d)$ is the next
dart counterclockwise around the black end of $d$, and $\sigma_1(d)$ the next dart around its white end. The
cycles of $\sigma_0\sigma_1$ (equivalently of $\sinf$, traversed in the opposite direction) are the boundary
walks of the faces~\citep{LandoZvonkin2004,JonesWolfart2016}. We call the pair of consecutive darts $(d,\sigma(d))$
at a vertex a \emph{corner}. Because each vertex carries a single cyclic order, a dessin always lives on an \emph{orientable} surface: ``counterclockwise'' must have the same meaning at every vertex, which is impossible on non-orientable surfaces such as the Möbius band, where embeddings require additional twist data~\citep{GrossTucker1987,MoharThomassen2001}. In the original setting, this is automatic, since Riemann surfaces are oriented by their complex structure and Belyi maps preserve orientation. \Cref{fig:dessin-example} shows an example: going around each face and recording which darts are passed gives the cycles of $\sinf$. The connection to algebraic geometry, which gives the theory its depth, is the following.

\begin{theorem}[Belyi; Grothendieck; see \citep{Belyi1979,Grothendieck1997Esquisse,LandoZvonkin2004}]
Connected dessins with $n$ darts, up to simultaneous conjugation of $(\sigma_0,\sigma_1)$, are in bijection with
isomorphism classes of pairs $(X,\beta)$, where $X$ is a compact Riemann surface and $\beta:X\to\mathbb{P}^1$
is a holomorphic map of degree $n$ unramified outside $\{0,1,\infty\}$. Such a surface $X$ is defined over
$\overline{\mathbb{Q}}$, and every curve defined over $\overline{\mathbb{Q}}$ arises in this way. The absolute
Galois group $\mathrm{Gal}(\overline{\mathbb{Q}}/\mathbb{Q})$ acts faithfully on dessins.
\end{theorem}

We use only the combinatorial side of the theory; under this correspondence the Riemann--Hurwitz formula becomes the Euler formula of \cref{prop:euler}, and the connection suggests future directions (\cref{sec:future}).

\subsection{Hypermaps: hypergraphs as dessins}
Walsh~\citep{Walsh1975} observed that a hypermap, that is, a hypergraph drawn on a surface, is the same thing
as a bipartite map: put a black vertex for every hypergraph vertex, a white vertex for every hyperedge, and an
edge for every incidence. Combined with the combinatorial definition above, this gives the dictionary we use
throughout the paper:

\begin{center}
\small
\begin{adjustbox}{max width=\linewidth}\begin{tabular}{@{}lll@{}}
\toprule
\textbf{Hypergraph} & \textbf{Dessin} & \textbf{Example: discussion forum}\\
\midrule
vertex $u$ & black vertex, a cycle of $\sigma_0$ & user\\
hyperedge $t$ & white vertex, a cycle of $\sigma_1$ & thread\\
incidence $(u,t)$ & dart & user's participation in a thread\\
order of the hyperedges at $u$ & cyclic order $\sigma_0$ at $u$ & order of a user's threads\\
order of the vertices in $t$ & cyclic order $\sigma_1$ at $t$ & order of a thread's participants\\
closed walk vertex--hyperedge--$\cdots$ & face, a cycle of $\sinf$ & loop of users via threads\\
\bottomrule
\end{tabular}\end{adjustbox}
\end{center}

The permutations $\sigma_0$, $\sigma_1$ and $\sinf$ play symmetric roles: the symmetric group $S_3$ acts on
dessins by permuting them (Jones--Singerman \emph{triality}~\citep{JonesSingerman1978}). For hypergraphs, the
transposition exchanging $\sigma_0$ and $\sigma_1$ (with inverses) is the familiar duality that exchanges
vertices and hyperedges. Triality says that faces are a third kind of cell, on equal footing with vertices and
hyperedges, that standard hypergraph representations do not have.

\subsection{Related work}
\paragraph{Higher-order networks and topology.}
Surveys of higher-order network models are given in~\citep{Battiston2020,Bick2023,Torres2021}.
Benson et al.~\citep{Benson2018} studied how groups close over time (\emph{simplicial closure}), which is
related in spirit to our cycle-closing events; temporal motifs were studied in~\citep{Paranjape2017}, partly
on Stack Exchange data. Topological data analysis of networks usually builds a simplicial or clique complex
and computes (persistent) homology~\citep{Carlsson2009,EdelsbrunnerHarer2010,Aktas2019}; sheaf-theoretic
methods attach vector spaces to incidences~\citep{HansenGhrist2019,Bodnar2022,aktas2025expert}; and topological deep learning
uses cell and combinatorial complexes~\citep{Hajij2022}. In all of these, the higher cells are either given by
the data or chosen by a filling rule. In a dessin, the 2-cells (faces) are \emph{forced} by the cyclic orders,
and they come with an Euler characteristic and a genus.

\paragraph{Topological graph theory.}
Embeddings of graphs in surfaces are described by rotation systems, going back to Heffter and Edmonds~\citep{Edmonds1960}; see~\citep{GrossTucker1987,MoharThomassen2001}. Computing the minimum genus of a graph is NP-hard~\citep{Thomassen1989}, while the maximum genus is computable in polynomial time~\citep{Xuong1979}. Random maps have nearly maximal genus~\citep{HarerZagier1986,PippengerSchleich2006,Gamburd2006}, which gives the reference point against which we measure local structure in data.

\paragraph{Hypergraph learning with incidences.}
Models that keep a separate state for each incidence~\citep{ChitraRaphael2019,Wang2023EDHNN} are close to the dart-based view; dessins add an order on incidences, a third type of cell (faces), and invariants such as the genus.

%=====================================================================================================
\section{Hypergraphs on surfaces: static theory}
\label{sec:static}

Throughout this section $\Hg=(U,T,\D)$ is a hypergraph and $(\sigma_0,\sigma_1)$ is a \emph{rotation system}
for it: $\sigma_0$ has one cycle per vertex, containing that vertex's darts, and $\sigma_1$ has one cycle per
hyperedge, containing that hyperedge's darts. We write $V=|U|$, $E=|T|$, $D=|\D|$, $F$ for the number of cycles of
$\sinf$, and $c$ for the number of components.

\subsection{Euler characteristic and genus}

\begin{proposition}[Euler formula for hypermaps]
\label{prop:euler}
There is an integer $g\ge 0$, the \emph{genus} of the dessin, such that
\begin{equation}
  V + E - D + F \;=\; 2c - 2g.
  \label{eq:euler}
\end{equation}
Here, the black and white vertices are the $V+E$ points of the decomposition, the $D$ darts are its edges, and the $F$ faces are its 2-cells. For a connected dessin, $g$ is the genus of the closed orientable surface on which the dessin is drawn; in
general $g$ is the sum of the genera of the components.
\end{proposition}

\begin{proof}
For each cycle $(d_1\,d_2\cdots d_k)$ of $\varphi=\sigma_0\sigma_1$, take a $2k$-gon whose sides, read counterclockwise, are labeled by the darts of the closed walk that alternates $\sigma_1(d_1),d_2,\sigma_1(d_2),d_3,\dots,$
$\sigma_1(d_k),d_1$; the darts $\sigma_1(d_i)$ are traversed from their white end to their black end and the darts $d_{i+1}$ from their black end to their white end. Every dart $e$ labels exactly two sides, once as $d_{i+1}$ in the face containing $e$ and once as $\sigma_1(d_i)$ in the face containing $\sigma_1^{-1}(e)$, and the two occurrences are traversed in opposite directions. Gluing the polygons along equally labeled sides therefore matches sides of opposite orientation, so the counterclockwise orientations of the polygons combine into an orientation of the result. Around a vertex with darts in cyclic order $(e_1\cdots e_m)$, the polygon corners between consecutive darts $e_j,e_{j+1}$ are glued in this cyclic order and form a single disk, because the rotation at each vertex is one cycle. Hence each component yields a closed orientable surface $S_i$ with a cell structure whose 0-cells are the black and white vertices, whose 1-cells are the darts, and whose 2-cells are the faces (see~\citep{JonesSingerman1978,MoharThomassen2001,LandoZvonkin2004}). Its Euler characteristic is $\chi(S_i)=V_i+E_i-D_i+F_i$, and by the classification of closed orientable surfaces $\chi(S_i)=2-2g_i$ with $g_i\ge0$. Summing over the $c$ components gives \eqref{eq:euler} with $g=\sum_i g_i$.
\end{proof}

In terms of the Belyi map, \eqref{eq:euler} is the Riemann--Hurwitz formula for a degree-$D$ cover of the
sphere branched over three points.

\subsection{The cycle decomposition}
Recall from \eqref{eq:beta1} that the cycle rank $\bone(\Hg)=D-V-E+c$ is the first Betti number of the incidence graph $B(\Hg)$, i.e., the number of independent cycles of the hypergraph. It depends only on the incidences, not on the rotation system. The next result shows how a rotation system splits these cycles into two kinds.
\begin{proposition}[Local and global cycles]
\label{prop:cycles}
For every rotation system,
\begin{equation}
  \bone(\Hg) \;=\; (F-c) \;+\; 2g.
  \label{eq:cycles}
\end{equation}
Homologically: let $S$ be the surface of the dessin. The inclusion of the incidence graph $B(\Hg)\hookrightarrow S$
induces a surjection $H_1(B;\mathbb{Z})\to H_1(S;\mathbb{Z})\cong\mathbb{Z}^{2g}$ whose kernel is spanned by
the boundaries of the faces and has rank $F-c$.
\end{proposition}

\begin{proof}
The first statement follows by substituting \eqref{eq:euler} into \eqref{eq:beta1}:
$\bone=D-V-E+c=(F-2c+2g)+c=(F-c)+2g$.
For the homological statement, consider the cellular chain complex of $S$,
$0\to C_2\xrightarrow{\partial_2}C_1\xrightarrow{\partial_1}C_0\to0$, where $C_2=\mathbb{Z}^F$ is spanned by
faces, $C_1=\mathbb{Z}^D$ by darts and $C_0=\mathbb{Z}^{V+E}$ by vertices. Since $C_1$ and $C_0$ and
$\partial_1$ are exactly the chain complex of the graph $B(\Hg)$, we have
$\ker\partial_1=H_1(B)\cong\mathbb{Z}^{\bone}$. Each component of $S$ is a closed orientable surface, so
$H_2(S)=\ker\partial_2\cong\mathbb{Z}^c$ and hence $\operatorname{rank}\operatorname{im}\partial_2=F-c$. Finally
$H_1(S)=\ker\partial_1/\operatorname{im}\partial_2\cong\mathbb{Z}^{2g}$
(see~\citep{Hatcher2002}), which shows that the map is onto with kernel $\operatorname{im}\partial_2$, the span of
the face boundaries.
\end{proof}

\Cref{fig:static} illustrates the proposition on the toy hypergraph of \cref{sec:static-example}, whose two independent cycles are fixed by the incidences ($\bone=2$). Under the planar rotation system of row 2 of \cref{tab:static-rotations} (\cref{fig:static-dessin}), each cycle bounds a face, so both are local: $F-c=2$ and $g=0$. Under the rotation system of row 3 (\cref{fig:static-torus}), which differs from row 2 only in the cyclic order at thread $a$, the dessin has a single face, and neither cycle bounds it: the two cycles wrap around the torus in two different directions; they cannot be shrunk to a point on the surface, and together they account for $2g=2$.

\Cref{prop:cycles} is the interpretive core of the paper. The cycle rank $\bone$ is a property of the
hypergraph alone: it counts how often vertices co-occur again (in a social setting, how often people ``meet
again''). The rotation system splits these cycles into two
kinds. \emph{Local} cycles bound faces: they are realized as small closed loops among vertices that are adjacent on
the surface. \emph{Global} cycles run around handles: they connect regions of the surface that are otherwise far
apart. The total never changes; only the split does. In Section \ref{sec:measures}, we will use this split to define topological features of the temporal hypergraphs. 

\subsection{Dependence on the rotation system}

A rotation system is chosen independently for each vertex and each hyperedge. A vertex $u$ has $\deg u$ darts and a hyperedge $t$ has $|t|$ darts, and $k$ darts can be arranged in $(k-1)!$ cyclic orders, so the number of rotation systems of $\Hg$ is
\begin{equation}
  \prod_{u\in U}(\deg u-1)!\;\prod_{t\in T}(|t|-1)!.
  \label{eq:nrot}
\end{equation}
For the small example of \cref{sec:static-example} below, \eqref{eq:nrot} gives $2!\cdot2!=4$ rotation systems; for the Data Science community studied in \cref{sec:results}, it gives about $10^{105{,}000}$. Different rotation systems give different surfaces. The smallest
attainable genus $\gamma_{\min}(\Hg)$ is the genus of the hypergraph's incidence graph; computing it is NP-hard
for graphs~\citep{Thomassen1989}. The largest attainable genus satisfies $\gamma_{\max}(\Hg)\le\lfloor\bone/2\rfloor$ (since
$F\ge c$ in \eqref{eq:cycles}) and can be computed in polynomial time thanks to Xuong's
characterization~\citep{Xuong1979}. Every rotation system gives a
genus in $[\gamma_{\min},\gamma_{\max}]$, and by \eqref{eq:cycles}, a smaller genus means more local cycles.

\begin{remark}[Random rotations]
\label{rem:random}
For uniformly random rotations, results on random gluings and random permutations show that the number of
faces of a large map grows only logarithmically in its size~\citep{HarerZagier1986,PippengerSchleich2006,
Gamburd2006}. By \eqref{eq:cycles}, the share of cycles that are local, $(F-c)/\bone$, then tends to zero. In
other words, a rotation system without structure produces an almost maximally tangled surface, and any
substantial local structure is a signature of organization. Similarly, by Dixon's theorem~\citep{Dixon1969}
two random permutations almost surely generate the full symmetric or alternating group, so the monodromy
group carries little information for large data dessins; we therefore work with faces and genus.
\end{remark}

\subsection{A first example: one hypergraph, two surfaces}
\label{sec:static-example}

\begin{figure}[!htbp]
\centering
\subfloat[]{\label{fig:static-sets}\includegraphics[width=0.27\linewidth]{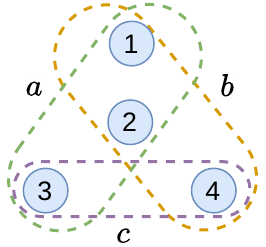}}\hfill
\subfloat[]{\label{fig:static-dessin}\includegraphics[width=0.27\linewidth]{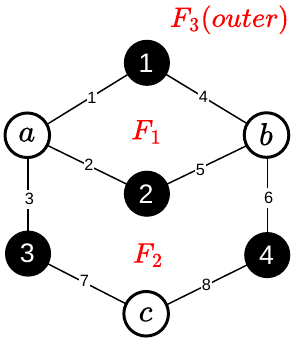}}\hfill
\subfloat[]{\label{fig:static-torus}\includegraphics[width=0.35\linewidth]{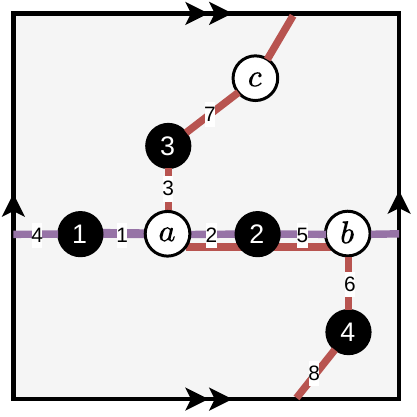}}
\caption{The toy hypergraph with hyperedges $a=\{1,2,3\}$, $b=\{1,2,4\}$, $c=\{3,4\}$, with $\bone=2$, and two of its dessins. (a) The hyperedges are drawn as sets. (b) Its dessin for the rotation system of row 2 of \cref{tab:static-rotations}, drawn in the plane. The eight darts are numbered. The rotations are read counterclockwise off the drawing, giving $\sigma_1=(1\,3\,2)(4\,5\,6)(7\,8)$ and faces $F_1=(2\,4)$ (users 1 and 2 meet twice), $F_2=(3\,8\,5)$ and the outer face $F_3=(1\,6\,7)$. Euler: $4+3-8+3=2$, so $g=0$; both independent cycles bound faces and are local. (c) The dessin for row 3 of \cref{tab:static-rotations}, $\sigma_1=(1\,2\,3)(4\,5\,6)(7\,8)$, which differs from row 2 only at thread $a$, drawn on a square whose opposite sides are glued. The purple cycle $1\,a\,2\,b$ leaves through the right side and returns through the left, and the red cycle $a\,3\,c\,4\,b\,2$ leaves through the top and returns through the bottom. There is a single face, so Euler gives $4+3-8+1=0$ and $g=1$; neither cycle bounds a face, so both are global, and together they are the two loops of the handle ($2g=2$).}
\label{fig:static}
\end{figure}

Consider the hypergraph with vertices $1,2,3,4$ and hyperedges (read them, for example, as users and threads) $  a=\{1,2,3\}, b=\{1,2,4\}, c=\{3,4\}$ (see \Cref{fig:static-sets}). It has $V=4$, $E=3$, $D=8$ darts and $c=1$ component, so $\bone=8-4-3+1=2$. Number the darts as $1=(1,a),\;2=(2,a),\;3=(3,a),\;4=(1,b),\;5=(2,b),\;6=(4,b),\;7=(3,c),\;8=(4,c).$ Every user lies in exactly two threads, so the rotation at each user is forced:
$\sigma_0=(1\,4)(2\,5)(3\,7)(6\,8)$. Thread $c$ has two darts and also a forced rotation $(7\,8)$. Threads $a$
and $b$ each have two possible cyclic orders, giving four rotation systems in total, in agreement
with~\eqref{eq:nrot}. \Cref{tab:static-rotations} lists them.

\begin{table}[t]
\centering
\caption{The four rotation systems of the hypergraph $a=\{1,2,3\}$, $b=\{1,2,4\}$, $c=\{3,4\}$, with
$\sigma_0=(1\,4)(2\,5)(3\,7)(6\,8)$ fixed. Faces are cycles of $\sinf=(\sigma_0\sigma_1)^{-1}$. All four have
$\bone=2$; two give a sphere and two give a torus. Rows 1 and 2 are mirror images of each other, and so are
rows 3 and 4. The cycles of $\sinf$ list each face's darts in the reverse of the order in which its boundary is walked; the cycles of $\varphi=\sigma_0\sigma_1$ (\cref{sec:online}) list them in the walking order.}
\label{tab:static-rotations}
\begin{adjustbox}{max width=\linewidth}\begin{tabular}{@{}cllccc@{}}
\toprule
 & $\sigma_1$ at $a$, $b$ & faces (cycles of $\sinf$) & $F$ & $g$ & $\bone=(F-c)+2g$ \\
\midrule
1 & $(1\,2\,3)(4\,6\,5)$ & $(1\,5)\,(2\,6\,7)\,(3\,8\,4)$ & 3 & 0 & $2+0$\\
2 & $(1\,3\,2)(4\,5\,6)$ & $(2\,4)\,(1\,6\,7)\,(3\,8\,5)$ & 3 & 0 & $2+0$\\
3 & $(1\,2\,3)(4\,5\,6)$ & $(1\,6\,7\,2\,4\,3\,8\,5)$ & 1 & 1 & $0+2$\\
4 & $(1\,3\,2)(4\,6\,5)$ & $(1\,5\,3\,8\,4\,2\,6\,7)$ & 1 & 1 & $0+2$\\
\bottomrule
\end{tabular}\end{adjustbox}
\end{table}

Take row 1. As maps, $\sigma_1$ sends $1\mapsto2\mapsto3\mapsto1$, $4\mapsto6\mapsto5\mapsto4$,
$7\leftrightarrow8$, and $\sigma_0$ swaps $1\leftrightarrow4$, $2\leftrightarrow5$, $3\leftrightarrow7$,
$6\leftrightarrow8$. Applying $\sigma_1$ first and then $\sigma_0$ gives
%\begin{table}[h!]
%\begin{center}
%\begin{adjustbox}{max width=\linewidth}\begin{tabular}{@{}cccc@{\qquad}cccc@{}}
%\toprule
%$d$ & $\sigma_1(d)$ & %$\sigma_0(\sigma_1(d))$ & & $d$ & %$\sigma_1(d)$ & $\sigma_0(\sigma_1(d))$ %& \\
%\midrule
%1 & 2 & 5 & & 2 & 3 & 7 & \\
%5 & 4 & 1 & back to 1 & 7 & 8 & 6 & \\
%3 & 1 & 4 & & 6 & 5 & 2 & back to 2 \\
%4 & 6 & 8 & & & & & \\
%8 & 7 & 3 & back to 3 & & & & \\
%\bottomrule
%\end{tabular}\end{adjustbox}
%\end{center}
%\end{table}
$\sigma_0\sigma_1=(1\,5)(2\,7\,6)(3\,4\,8)$ and $\sinf$ has the same cycles traversed backwards. Each step
``apply $\sigma_1$, then $\sigma_0$'' walks along the boundary of a face: arrive at a thread, turn to the next
participant, walk to that user, turn to the user's next thread. The face $(2\,7\,6)$, for instance, is the
closed walk $2\to a\to3\to c\to4\to b\to2$, a triangle of users connected through three different threads.

In rows 1 and 2 (the sphere), the two-dart face is the loop $1\to a\to2\to b\to1$ (users 1 and 2 meet in two threads), and the two longer faces lie on either side of the path $a\to3\to c\to4\to b$: one closes it through user~2 and the other through user~1. In rows 3 and 4 (the torus),
no cycle is closed as a face: there is one face of length 8, and both cycles are global. The same hypergraph and
the same $\bone=2$ lead to two different surfaces, decided entirely by the order of the participants of
threads $a$ and $b$. \Cref{fig:static} draws the sphere of row 2 and the torus of row 3. \Cref{sec:simplicial} compares these surfaces with the simplicial completion of the hypergraph.

%=====================================================================================================
%=====================================================================================================
\subsection{Relation to the simplicial completion}
\label{sec:simplicial}

The ``topology of a hypergraph'' usually means the homology of the simplicial complex $\Delta(\Hg)$ generated by
its hyperedges (each hyperedge is filled in as a simplex, together with all its faces).

\begin{proposition}
\label{prop:simplicial}
$H_1(\Delta(\Hg))$ is a quotient of $H_1(B(\Hg))$. Hence both $H_1(\Delta(\Hg))$ and $H_1(S)$, where $S$ is the surface of the dessin of any rotation system (\cref{prop:cycles}), are quotients of the same group $H_1(B(\Hg))\cong\mathbb{Z}^{\bone}$.
\end{proposition}

\begin{proof}
Map $B(\Hg)$ into $\Delta(\Hg)$ by sending a user $u$ to the vertex $u$, a thread $t$ to the barycenter of the
simplex $t$, and the dart $(u,t)$ to the straight segment between them inside the simplex $t$. Every edge
$\{u,v\}$ of $\Delta(\Hg)$ lies in some simplex $t\ni u,v$ and is homotopic, inside that simplex, to the path
$u\to t\to v$ in the image of $B(\Hg)$. Hence every loop in the 1-skeleton of $\Delta(\Hg)$, and therefore every
class of $H_1(\Delta(\Hg))$, is homologous to the image of a cycle of $B(\Hg)$, so the induced map on $H_1$ is
onto.
\end{proof}

The two quotients need not agree. In the example of \cref{sec:static-example}, $\Delta(\Hg)$ has $\bone(\Delta)=1$: the loop
$2\to3\to4\to2$ is a genuine hole because the edge $\{3,4\}$ belongs to no filled triangle, while the loop
through users 1 and 2 is filled by the triangles of $a$ and $b$. The sphere of rows 1--2 fills \emph{both}
cycles of $B(\Hg)$ (the second one by the face bounded by $2\to a\to3\to c\to4\to b\to2$), so $H_1(S)=0$; the torus of
rows 3--4 fills neither. The dessin is therefore not a way of computing the homology of $\Delta(\Hg)$. It is a
different, order-sensitive way of deciding which cycles of the hypergraph are ``small''.

\subsection{Symmetries}
\label{sec:symmetries}
Relabeling the darts by a permutation $\pi$ conjugates both rotations,
$(\sigma_0,\sigma_1)\mapsto(\pi\sigma_0\pi^{-1},\pi\sigma_1\pi^{-1})$, and changes none of the quantities above.
All our measures are therefore invariant under relabeling. Reflecting the surface (reversing all rotations)
also leaves $F$, $c$ and $g$ unchanged; this is why the rows of \cref{tab:static-rotations} come in mirror
pairs.

%=====================================================================================================
\section{Growing dessins for temporal hypergraphs}
\label{sec:online}
 
We now let the hypergraph grow. The input is a temporal hypergraph given as a time-ordered sequence of
\emph{events} $e_k=(u_k,t_k,a_k)$, $k=1,\dots,D$: at step $k$ vertex $u_k$ becomes incident to hyperedge $t_k$
for the first time. The optional \emph{anchor} $a_k$ is an earlier incidence of the same hyperedge $t_k$ to
which the new incidence is related, or $a_k=\varnothing$ if there is none (for example when $t_k$ is new). In a
forum, $a_k$ is the incidence of the author of the post that $u_k$ replies to; \cref{sec:domains}
gives other examples. After $k$ events, we have a hypergraph $\Hg_k$ with $k$ darts. Ties in time can be broken by any fixed
rule. Repeated incidences (a vertex that appears in the same hyperedge again) create no new event, because the
incidence already exists.

As explained in \cref{sec:intro}, we do not read the rotations directly off the timestamps (\cref{sec:toy,sec:negative} show why): the data decide which incidence arrives when, and where available its anchor, while the position of each new dart is chosen from the faces of the surface built so far.

\subsection{Inserting a dart}
 
The growing dessin maintains a rotation system for $\Hg_k$. Event $k$ adds one dart $z=(u_k,t_k)$:
\begin{itemize}[leftmargin=*]
  \item at the vertex if $u_k$ is new, $z$ forms its own cycle, $\sigma_0'(z)=z$. Otherwise, we choose a dart $x$   of $u_k$ and insert $z$ into the corner after it: $x\mapsto z\mapsto\sigma_0(x)$.
  \item at the hyperedge if $t_k$ is new, $\sigma_1'(z)=z$. Otherwise, we choose a dart $y$ of $t_k$ and insert $z$ into the corner after it: $y\mapsto z\mapsto\sigma_1(y)$.
\end{itemize}
Nothing else changes. In particular, \emph{existing rotations are never rearranged}: the rotation system at any
moment is the accumulated result of the whole insertion history. Time decides the order of insertions; the
\emph{insertion rule} (\cref{sec:rules}) decides the corners $x$ and $y$.
 
\subsection{Event classification}
Each insertion adds one dart and may change the number of faces, components, the cycle rank, and the genus. We now show that these changes are completely determined by a single local question: whether the two corners chosen for the new dart lie on the same face. Since the faces are the cycles of $\varphi=\sigma_0\sigma_1$, and inserting a dart changes $\varphi$ essentially by a transposition, the argument rests on a standard fact about how multiplying a permutation by a transposition changes its number of cycles.
 
\begin{lemma}[Transposition lemma]
\label{lem:transposition}
Let $\pi$ be a permutation of a finite set and $p\ne q$. If $p$ and $q$ lie in the same cycle of $\pi$, then
$\pi\circ(p\;q)$ has exactly one more cycle than $\pi$; otherwise it has exactly one fewer.
\end{lemma}
 
\begin{proof}
Write the cycle(s) of $\pi$ containing $p$ and $q$. If both are in one cycle
$(p\;a_1\cdots a_r\;q\;b_1\cdots b_s)$, then $\pi\circ(p\,q)$ maps $p\mapsto\pi(q)=b_1$ and $q\mapsto\pi(p)=a_1$,
producing the two cycles $(p\;b_1\cdots b_s)$ and $(q\;a_1\cdots a_r)$. If they lie in different cycles, the same computation read backward joins them into one. All other cycles are unchanged.
\end{proof}
 
The hyperedge corner after $y$ lies on the face containing $y$, and the vertex corner after $x$ lies on the face containing $w=\sigma_1^{-1}(x)$, because
$\varphi(w)=\sigma_0(x)$ is the dart that follows that corner.
 
\begin{theorem}[Event classification]
\label{thm:events}
Every insertion is of exactly one of the following five types, with the stated changes in the number of
faces $F$, components $c$, cycle rank $\bone$ and genus $g$:
\begin{center}
\begin{adjustbox}{max width=\linewidth}\begin{tabular}{@{}lllcccc@{}}
\toprule
type & vertex & hyperedge & $\Delta F$ & $\Delta c$ & $\Delta\bone$ & $\Delta g$ \\
\midrule
\emph{new} & new & new & $+1$ & $+1$ & $0$ & $0$\\
\emph{pendant} & \multicolumn{2}{l}{exactly one of them new} & $0$ & $0$ & $0$ & $0$\\
\emph{bridge} & old & old, different components & $-1$ & $-1$ & $0$ & $0$\\
\emph{local} & old & old, same component, corners on the same face & $+1$ & $0$ & $+1$ & $0$\\
\emph{handle} & old & old, same component, corners on different faces & $-1$ & $0$ & $+1$ & $+1$\\
\bottomrule
\end{tabular}\end{adjustbox}
\end{center}
\end{theorem}
 
\begin{proof}
We compute $\varphi'=\sigma_0'\sigma_1'$ from $\varphi$.
 
\emph{Both endpoints new.} Then $\sigma_0'(z)=\sigma_1'(z)=z$, so $\varphi'(z)=z$ is a new face, and $z$ forms
a new component. $V$, $E$ and $D$ each increase by one, so by \eqref{eq:euler} $\Delta(2c-2g)=1+1-1+1=2$, and
with $\Delta c=1$ we get $\Delta g=0$; $\Delta\bone=1-1-1+1=0$ by \eqref{eq:beta1}.
 
\emph{New vertex, old hyperedge.} Then $\sigma_0'(z)=z$ and $y\mapsto z\mapsto\sigma_1(y)$. We get
$\varphi'(y)=\sigma_0'(z)=z$ and $\varphi'(z)=\sigma_0(\sigma_1(y))=\varphi(y)$; all other values are unchanged.
So $z$ is spliced onto the face of $y$, and $F$ remains unchanged. $V$ and $D$ increase by one, $c$ is unchanged, so
$\Delta g=0$ and $\Delta\bone=0$. The case of a new hyperedge and an old vertex is symmetric:
$\varphi'(w)=z$ and $\varphi'(z)=\varphi(w)$, with $w=\sigma_1^{-1}(x)$.
 
\emph{Both endpoints old.} Since $(u_k,t_k)$ is a new incidence, $x$ is not a dart of $t_k$, so
$w=\sigma_1^{-1}(x)\ne y$. The only changed values are
\[
  \varphi'(y)=\sigma_0'(z)=\sigma_0(x)=\varphi(w),\qquad
  \varphi'(w)=\sigma_0'(x)=z,  \qquad \varphi'(z)=\sigma_0'(\sigma_1(y))=\varphi(y).
\]
Contracting $z$ out of its cycle (replacing $w\mapsto z\mapsto\varphi(y)$ by $w\mapsto\varphi(y)$) gives
the permutation $y\mapsto\varphi(w)$, $w\mapsto\varphi(y)$, $d\mapsto\varphi(d)$ otherwise, which is exactly
$\varphi\circ(y\;w)$; contracting an element out of a cycle of length at least two does not change the number of
cycles. By \cref{lem:transposition}, $F$ increases by one if $y$ and $w$, that is, the two chosen corners, lie
on the same face, and decreases by one otherwise. $V$ and $E$ are unchanged and $D$ increases by one.
 
If $u_k$ and $t_k$ lie in different components, they cannot share a face, so $\Delta F=-1$, $\Delta c=-1$,
$\Delta(2c-2g)=-1-1=-2$, hence $\Delta g=0$; and $\Delta\bone=1-1=0$ (a bridge). If they lie in the same
component, $\Delta c=0$ and $\Delta\bone=+1$. Then either $\Delta F=+1$, so $\Delta(2c-2g)=0$ and $\Delta g=0$
(local), or $\Delta F=-1$, so $\Delta(2c-2g)=-2$ and $\Delta g=+1$ (handle).
\end{proof}
 
Geometrically, a local event draws a new edge across a face (a disk) between two points on its boundary, which
cuts the disk into two disks. A handle event connects two different faces; to draw the edge without crossings,
the surface needs a new handle, and the two faces merge into one.
 
\begin{corollary}[Conservation law]
\label{cor:conservation}
Let $L_k$ and $H_k$ be the numbers of local and handle events among the first $k$ events. Then at every time $k$
\[
  \bone(\Hg_k)=L_k+H_k,\qquad g_k=H_k,\qquad F_k-c_k=L_k-H_k.
\]
In particular $L_k\ge H_k$ at all times, and the share of local events among cycle-closing events is at least
$1/2$.
\end{corollary}
 
\begin{proof}
Sum the changes in \cref{thm:events} over the first $k$ events, starting from the empty dessin. Only local and
handle events change $\bone$ (by $+1$ each), only handle events change $g$, and $F-c$ changes by $+1$ for local,
$-1$ for handle and $0$ for the other three types. Since every component has at least one face, $F_k\ge c_k$.
\end{proof}
 
\begin{remark}
Whether an event closes a cycle (local or handle) depends only on the hypergraph: it does so exactly when the
vertex and the hyperedge already exist and are already connected. Whether it closes the cycle \emph{locally}
depends on the rotation system built so far, that is, on the history and on the insertion rule.
\end{remark}
 
\subsection{Insertion rules}
\label{sec:rules}
 
An insertion rule chooses the corners $x$ and $y$. For a cycle-closing event, this choice decides whether the event is local or a handle (\cref{thm:events}); for pendant and bridge events, it changes no counts but decides on which faces the new dart lies, and therefore which later events can close locally, as step~6 of the worked example shows. Cycle-closing events are therefore handled by a shared-face search, and all other events by one of three placement rules.
 
\begin{description}[leftmargin=*]
  \item[Shared-face search (cycle-closing events).] If the event closes a cycle and some face contains both a
  corner of $u_k$ and a corner of $t_k$, insert the dart there; the event is then local. If several such faces exist, we pick one uniformly at random, and within it a corner of each endpoint at random. If there is none,
  the event is necessarily a handle and the corners are chosen as for pendant events. By
  \cref{thm:events}, this choice maximizes $F$ after the current step; it is a \emph{myopically optimal}
  rule. It is not globally optimal, as the worked example in \cref{sec:toy} shows.
  \item[Anchor rule (main rule).] For pendant and bridge events, insert at the hyperedge right after the
  anchor $a_k$, falling back to the hyperedge's first dart if there is no anchor; and insert at the vertex right
  after the vertex's most recent hyperedge. The assumption behind this rule is that a new incidence joins the
  neighborhood of the incidence it is related to. In our application the anchor is given by the reply
  structure, and we call this instance the \emph{reply rule}.
  \item[Hub rule.] As the anchor rule, but at the hyperedge, insert after the dart of the vertex with the largest
  number of incidences so far. The assumption is that new incidences attach to the most central vertex of the
  hyperedge. This rule needs no anchors, so it applies to any temporal hypergraph.
  \item[Random rule.] Insert into uniformly random corners. This is the floor: growth without structure, apart
  from the shared-face search.
\end{description}
 
The shared-face search is used by all three rules. We also report the \emph{shuffled ablation}, in which the
same events are processed in a uniformly random order (so the final hypergraph is identical) with the random
rule, and an \emph{offline ceiling} (\cref{sec:ceiling}). The construction requires only a time-ordered sequence of incidences: anchors are optional, since the hub and random rules require none, and \cref{tab:domains} lists natural anchors for e-mail, co-authorship, meetings, messaging groups, and reaction networks.
 
\subsection{Algorithm and complexity}
\label{sec:algorithm}
 
\begin{algorithm}[t]
\caption{Growing dessin: insertion of event $(u,t,a)$}
\label{alg:online}
\begin{algorithmic}[1]
\State $z\gets$ new dart
\If{$u$ and $t$ both exist and are in the same component}
  \State $S\gets\{\text{faces with a corner of }u\}\cap\{\text{faces with a corner of }t\}$
  \If{$S\ne\emptyset$} \State choose $f\in S$; choose $x$ (corner of $u$ on $f$) and $y$ (corner of $t$ on $f$)
  \Else \State choose $x,y$ by the insertion rule \EndIf
\Else \State choose $x$ (if $u$ exists) and $y$ (if $t$ exists) by the insertion rule
\EndIf
\State update $\sigma_0$: $x\mapsto z\mapsto\sigma_0(x)$ (or $\sigma_0(z)=z$); update $\sigma_1$ likewise
\State update $\varphi$ at the at most four darts $z$, $y$, $\sigma_1^{-1}(x)$, $\sigma_1^{-1}(z)$
\State retrace the affected faces in lockstep; relabel only the pieces that finish first
\State classify the event by $\Delta F$ and $\Delta c$ (\cref{thm:events}) and update $L$, $H$
\end{algorithmic}
\end{algorithm}
 
We store $\sigma_0$, $\sigma_1$, $\sigma_1^{-1}$ and $\varphi$ as arrays, a face label for every dart and the set of darts of each face, and a union--find structure for components. The inverse $\sigma_1^{-1}$ is needed, and $\sigma_0^{-1}$ is not, because $\varphi=\sigma_0\sigma_1$ applies $\sigma_1$ first: inserting a dart $z$ after $x$ at a vertex and after $y$ at a hyperedge changes $\varphi$ only at $z$, $y$, $\sigma_1^{-1}(x)$ and $\sigma_1^{-1}(z)$, and the corner $(x,\sigma_0(x))$ lies on the face of $\sigma_1^{-1}(x)$, since $\varphi(\sigma_1^{-1}(x))=\sigma_0(x)$. The shared-face test costs
$O(\deg u+|t|)$ label lookups. After an insertion only the faces through the affected darts can change. We
trace these faces simultaneously, one step at a time, and stop as soon as all but one have closed; only the
closed pieces receive new labels, and when two faces merge, the smaller one is relabeled into the larger (\cref{sec:sm-lockstep}).
Hence the work per event is proportional to the size of the smaller affected piece rather than to the size of
the largest face, which matters because data dessins develop very large faces. In our pure Python
implementation, a full run takes about 15 seconds for Data Science ($99{,}140$ events) and about two minutes
for Academia ($308{,}289$ events) on one CPU core.
 
Randomized tests confirm that after every insertion in thousands of random event sequences, the maintained faces agree with a full recount of the cycles of $\varphi$, and the identities in \cref{cor:conservation} hold exactly.
 
\subsection{Measures}
\label{sec:measures}
 
The raw counts $F$, $g$, and $\bone$ all increase with the size of the hypergraph, so they cannot be compared directly across hypergraphs or over time. We therefore summarize a growing dessin by normalized quantities that answer two separate questions: how often a new incidence closes a cycle at all, which depends only on the hypergraph, and how the cycles that are closed split between local cycles ($F-c$) and global ones ($2g$), which depends on the surface. By \cref{prop:cycles}, these two questions together account for all of $\bone$.
 
\begin{definition}[Measures of a growing dessin]
At any time, with $\bone>0$:
\begin{enumerate}[leftmargin=*]
  \item the \emph{cycle rate} $\rho=\bone/D$, the share of incidences that closed a cycle (order-free);
  \item the \emph{handle fraction} $\hfrac=2g/\bone\in[0,1]$, the share of cycles that became global;
  \item the \emph{local cycle percentage} $\nloc=100\,(F-c)/\bone=100\,(1-\hfrac)$, the percentage of independent cycles that are local, that is, that bound faces;
  \item the \emph{local share} $s=L/(L+H)=\tfrac12(1+\nloc/100)\in[\tfrac12,1]$;
  \item the \emph{event composition}: the shares of new, pendant, bridge, local and handle events among all
  events.
\end{enumerate}
\end{definition}
 
By \cref{cor:conservation}, $\hfrac$, $\nloc$ and $s$ are three expressions of the same quantity; we report $\nloc$ when differences are small and $\hfrac$ when the genus itself is the point. The cycle rate is a property of the hypergraph and the handle fraction a property of the surface; the event composition shows both, since the local and handle shares add up to $\rho$ and their ratio determines $\hfrac$. A handle fraction close to $1$ means that vertices that co-occur again almost always do so across distant regions of the surface built so far; in a social setting, the question is whether people who meet again do so among neighbors or across the community. Structureless growth drives $\hfrac$ towards $1$ (\cref{rem:random}), so the distance from $1$ is the signal. Like the face count of random maps, the local share decreases as a dessin grows, so hypergraphs of different sizes must be compared at the same number of cycles: we report every measure as a function of $\bone$ and compare communities at matched values $\bone\in\{5{,}000,\,10{,}000,\,20{,}000\}$.

\subsection{Offline ceiling: face maximization}
\label{sec:ceiling}
To bracket the online results, we also compute, for a fixed hypergraph, a rotation system with as many faces as we can find, that is, a heuristic minimum-genus embedding. Starting from random rotations, we repeatedly move one dart to another corner of the same vertex. Writing the move at a hyperedge as $\sigma_1'=\sigma_1\circ(p\;x\;y)$ with $p=\sigma_1^{-1}(x)$, we get $\varphi'=\varphi\circ(p\;y)\circ(p\;x)$, so by \cref{lem:transposition} the change in $F$ is $-2$, $0$ or $+2$ and can be read in constant time (\cref{sec:sm-moves}). We run simulated annealing~\citep{Kirkpatrick1983} followed by hill climbing. On 40 small hypergraphs whose optimum we found by enumeration, the heuristic always found it; on large samples, the search has not fully converged, so the ceiling is a \emph{lower bound} on the maximum number of faces. Because it uses the final hypergraph, it tells us how planar a hypergraph \emph{could} be, in hindsight.
 
%=====================================================================================================
\subsection{Instantiating the construction in other domains}
\label{sec:domains}
 
The construction needs only a time-ordered sequence of incidences. Anchors are optional: without them, the
hub rule or the random rule can be used, and the shared-face search works in all cases.
\Cref{tab:domains} lists several temporal hypergraphs and natural choices of anchors.
 
\begin{table}[h]
\centering
\caption{Examples of temporal hypergraphs and how the growing dessin applies to them.}
\label{tab:domains}
\small
\begin{adjustbox}{max width=\linewidth}\begin{tabular}{@{}>{\raggedright\arraybackslash}p{2.7cm}>{\raggedright\arraybackslash}p{2.0cm}>{\raggedright\arraybackslash}p{2.5cm}>{\raggedright\arraybackslash}p{3.2cm}>{\raggedright\arraybackslash}p{4.3cm}@{}}
\toprule
domain & vertices & hyperedges & time of an incidence & natural anchor\\
\midrule
discussion forums (this paper) & users & threads & first post of the user in the thread & author of the post
being replied to\\
e-mail & people & conversation threads & first message sent or received in the thread & sender of the message
that added the person\\
co-authorship & authors & papers, or research projects & publication date (with author order as a tie-break)
& preceding author in the author list, or the corresponding author\\
meetings, events & participants & meetings & time of joining or invitation & the person who invited them\\
messaging groups & users & group chats & time of joining the group & the member who added them\\
reaction networks & chemical species & reactions & discovery or publication date & a species already known to
take part in the reaction\\
\bottomrule
\end{tabular}\end{adjustbox}
\end{table}
 
In each case the construction answers the same question: as the hypergraph grows, does each new cycle close
among vertices that are already adjacent on the surface (a local event), or does it join distant regions (a
handle event)? The measures of \cref{sec:measures} are defined identically in every domain; only the
interpretation of ``meeting again'' changes.

\subsection{A worked example, step by step}
\label{sec:toy}
 
We now follow the growing dessin on the hypergraph of \cref{sec:static-example}. To make the anchors concrete,
we tell its history as a discussion forum (vertices are users, hyperedges are threads, and the anchor of an
answer is the incidence of the user it replies to), with the following eight events. User~1 asks question~$a$; users~2 and~3 answer it; user~1 asks question~$b$;
users~2 and~4 answer it; user~3 asks question~$c$; and user~4 answers it. All answers to $a$ and $b$ reply to
their asker, user~1, and user~4's answer to $c$ replies to user~3. We use the reply rule with the shared-face
search. \Cref{tab:toy-steps} lists every step: the event, its type, the rotations that change, and the counts $V$, $E$, $D$, $F$, $c$, $\bone$ and $g$, so that the identity $\bone=(F-c)+2g$ can be checked after each step. %\Cref{fig:toy-steps-a,fig:toy-steps-b} draw every step; the rotations, the event type and the four counts $F$, $c$, $\bone$, $g$ are printed in each panel, and \cref{tab:toy-steps} collects them. In the drawings, rotations are read counterclockwise and agree with the rotations maintained by the algorithm; we checked this automatically for every panel.
 
%\begin{figure}[!htbp]
%\centering
%\includegraphics[width=0.92\linewidth]{toy_steps_a.pdf}
%\caption{The growing dessin of the toy history, steps 1--4 (reply rule). Black: users; white: threads. At the
%top of each panel: the rotation of every vertex with at least two darts (dart $2a$ means user 2 in thread
%$a$), read counterclockwise. At the bottom: event type and counts, with the check $\bone=(F-c)+2g$. The
%drawing is a tree throughout: one face, no cycles.}
%\label{fig:toy-steps-a}
%\end{figure}
 
%\begin{figure}[!htbp]
%\centering
%\includegraphics[width=0.92\linewidth]{toy_steps_b.pdf}
%\caption{Steps 5--8 of the toy history. In step 5 the first cycle closes locally: the loop $1$--$a$--$2$--$b$
%bounds face A (shaded); everything else is face B. In step 6 the reply rule places user 4, who replies to user
%1, inside face A. In step 8 user 4 (on face A only) answers thread $c$ (on face B only), so the new edge must
%leave face A: the two faces merge and a handle is created ($g=1$). The last panel cannot be drawn in the plane
%without a crossing; the dashed edge passes over the new handle.}
%\label{fig:toy-steps-b}
%\end{figure}
 
\begin{table}[t]
\centering
\caption{All numbers of the toy history (reply rule).
Rotations are listed only where they change.}
\label{tab:toy-steps}
\small
\begin{adjustbox}{max width=\linewidth}\begin{tabular}{@{}clllccccccc@{}}
\toprule
step & event & type & rotation change & $V$ & $E$ & $D$ & $F$ & $c$ & $\bone$ & $g$\\
\midrule
1 & 1 asks $a$ & new & $1{:}(1a)$, $a{:}(1a)$ & 1 & 1 & 1 & 1 & 1 & 0 & 0\\
2 & 2 answers $a$ & pendant & $a{:}(1a\;2a)$ & 2 & 1 & 2 & 1 & 1 & 0 & 0\\
3 & 3 answers $a$ & pendant & $a{:}(1a\;3a\;2a)$ & 3 & 1 & 3 & 1 & 1 & 0 & 0\\
4 & 1 asks $b$ & pendant & $1{:}(1a\;1b)$ & 3 & 2 & 4 & 1 & 1 & 0 & 0\\
5 & 2 answers $b$ & local & $2{:}(2a\;2b)$, $b{:}(1b\;2b)$ & 3 & 2 & 5 & 2 & 1 & 1 & 0\\
6 & 4 answers $b$ & pendant & $b{:}(1b\;4b\;2b)$ & 4 & 2 & 6 & 2 & 1 & 1 & 0\\
7 & 3 asks $c$ & pendant & $3{:}(3a\;3c)$ & 4 & 3 & 7 & 2 & 1 & 1 & 0\\
8 & 4 answers $c$ & handle & $4{:}(4b\;4c)$, $c{:}(3c\;4c)$ & 4 & 3 & 8 & 1 & 1 & 2 & 1\\
\bottomrule
\end{tabular}\end{adjustbox}
\end{table}
 
\paragraph{Steps 1--4: growing a tree.}
Step 1 creates user~1, thread~$a$ and the dart $1a$: a single edge on a sphere, with one face ($1+1-1+1=2$, so $g=0$). In steps 2 and 3, users~2 and~3 are new, and the reply rule inserts their darts at $a$ right after $1a$, the dart of the author they reply to, giving the rotation $(1a\;3a\;2a)$. In step~4 thread~$b$ is new, and at user~1 its dart goes after the user's most recent thread, giving $(1a\;1b)$. All four events are pendant: the drawing is a tree with a single face and $\bone=0$.

\paragraph{Step 5: the first cycle closes locally.}
User~2 answers $b$. Both user~2 and thread~$b$ exist and are connected (through user~1), so this event closes
a cycle, $1\to a\to2\to b\to1$, and $\bone$ becomes $1$. The only face of the tree contains corners of both
user~2 and thread~$b$, so the shared-face search inserts the dart there. By \cref{thm:events}, the face splits
into two: face A, bounded by the new loop, and face B, everything else (user~3 hangs off thread $a$ in face B).
Now $F=2$, $c=1$, $g=0$, and the check reads $\bone=1=(2-1)+0$. One local event, no handle events.
 
\paragraph{Step 6: a placement decision.}
User~4 is new and answers $b$, replying to user~1. The reply rule inserts the dart at $b$ right after $1b$,
giving the rotation $(1b\;4b\;2b)$. The corner between $1b$ and $2b$ on that side lies inside face~A, so user~4
is placed \emph{inside} the loop of users 1 and 2. This is a pendant event and changes no counts, but it decides
the future. 
 
\paragraph{Step 7}
User~3 asks question $c$; its dart at user~3 goes after $3a$, so thread~$c$ also lies on face~B. Pendant.

\paragraph{Step 8: the second cycle becomes a handle.}
User~4 answers $c$. Both exist and are connected, so the event closes the cycle
$4\to c\to3\to a\to1\to b\to4$ and $\bone$ becomes $2$. The shared-face search now finds no face containing
corners of both: user~4 has a single dart, whose corners lie on face~A, and thread~$c$ has a single dart, whose
corners lie on face~B. By \cref{thm:events} the two faces merge and the genus increases: $F=1$, $c=1$, $g=1$,
and the check reads $\bone=2=(1-1)+2\cdot1$. The final surface is a torus.
 
\paragraph{The alternative history}
Suppose that in step~6 user~4's dart had been inserted after $2b$ instead, giving the rotation
$(1b\;2b\;4b)$. The corner after $2b$ lies on face~B, so user~4 would sit outside the
loop, on the same face as user~3. In step~8, user~4 and thread~$c$ would then share face~B, the new edge would
split it, and the event would be local: $F=3$, $c=1$, $g=0$, and $\bone=2=(3-1)+0$. The final surface would be a
sphere with the three faces of row 2 of \cref{tab:static-rotations}. \Cref{tab:toy-measures} compares the two histories.
 
%\begin{figure}[!htbp]
%\centering
%\includegraphics[width=0.85\linewidth]{toy_alternative.pdf}
%\caption{The alternative history: in step $6'$ user 4 is inserted after $2b$, which places it in face B
%(outside the shaded loop of users 1 and 2). In step $8'$ user 4 and thread $c$ then share a face, the event is
%local, and the final surface is a sphere with three faces.}
%\label{fig:toy-alt}
%\end{figure}
 
\begin{table}[t]
\centering
\caption{Measures of the actual (reply rule) and alternative toy histories. Same hypergraph, same cycles,
different surfaces.}
\label{tab:toy-measures}
\begin{adjustbox}{max width=\linewidth}\begin{tabular}{@{}lcc@{}}
\toprule
 & reply rule (actual) & alternative placement\\
\midrule
events (new / pendant / bridge / local / handle) & 1 / 5 / 0 / 1 / 1 & 1 / 5 / 0 / 2 / 0\\
cycle rank $\bone$ & 2 & 2\\
cycle rate $\rho=\bone/D$ & $2/8=0.25$ & $2/8=0.25$\\
genus $g$ & 1 & 0\\
handle fraction $\hfrac=2g/\bone$ & 1.0 & 0.0\\
local cycle percentage $\nloc$ & 0 & 100\\
local share $s=L/(L+H)$ & 0.5 & 1.0\\
\bottomrule
\end{tabular}\end{adjustbox}
\end{table}
 
The cycle rate, a property of the hypergraph, is the same in both histories, while the handle fraction, the genus, and the faces, properties of the surface, differ. The outcome is decided by a single placement of a newcomer three steps before the cycle closes, and the shared-face search cannot repair it, because user~4 and thread~$c$ each had only one corner to choose from. The insertion rule is therefore part of the method's definition, much as filtration is part of a persistent homology analysis; it is a modeling assumption rather than a guarantee, so we test the conclusions under several rules.

The same example also shows why cyclic orders read directly off the timestamps fail. Face~A, $(1a\;2b)$, exists only if, among other conditions, the dart after $2b$ in thread $b$ is $1b$. Users joined $b$ in the order $1,2,4$, so the timestamp order would put $4b$ after $2b$, and the loop of users~1 and~2 would not bound a face, merely because user~4 happened to join $b$ after user~2. The growing dessin instead placed user~4 next to its anchor (step~6), giving the rotation $(1b\;4b\;2b)$, which keeps face~A even though it is not the time order. A face of length $k$ requires about $k$ such conditions at once; when each holds only by chance, long faces almost never close, which is why timestamp rotations behave like random ones (\cref{sec:negative}).
 
%=====================================================================================================
%=====================================================================================================
\section{Data and experimental setup}
\label{sec:setup}

%\subsection{Communities}
We use the public Stack Exchange data~\citep{StackExchangeDump} for five communities chosen to span
different topics and interaction cultures: two technical communities (Data Science and Artificial
Intelligence), two natural-science communities (Biology and Astronomy) and one discussion-oriented community
about academic life (Academia). All data end on 31 March 2024. \Cref{tab:profile} summarizes them.

\begin{table}[t]
\centering
\caption{Order-free profile of the five communities. A participation is a user's first post (question,
answer, or comment) in a thread. ``Repeat pairs'' is the share of co-participating user pairs that meet in at
least two threads; ``top 1\%'' is the share of all participations made by the 1\% most active users; ``reply''
is the share of participations that reply to someone else's post. $\rho=\bone/D$ is the cycle rate of the final
hypergraph.}
\label{tab:profile}
\small
\begin{adjustbox}{max width=\linewidth}\begin{tabular}{@{}lrrrrr@{}}
\toprule
 & Data Science & Artif.\ Intelligence & Biology & Astronomy & Academia\\
\midrule
first post & 2014-05 & 2016-08 & 2011-05 & 2013-09 & 2011-10\\
threads $E$ & 36{,}766 & 12{,}372 & 28{,}780 & 14{,}624 & 43{,}784\\
users $V$ & 29{,}309 & 9{,}509 & 18{,}194 & 9{,}706 & 35{,}947\\
participations $D$ & 99{,}140 & 32{,}727 & 101{,}339 & 64{,}720 & 308{,}289\\
participants per thread & 2.70 & 2.65 & 3.52 & 4.43 & 7.04\\
threads with $\ge3$ participants & 43.8\% & 43.8\% & 65.3\% & 78.1\% & 89.0\%\\
users in a single thread & 69.5\% & 69.5\% & 68.7\% & 65.9\% & 65.7\%\\
repeat pairs & 6.4\% & 9.2\% & 14.1\% & 15.9\% & 18.6\%\\
top 1\% share & 38.8\% & 40.2\% & 54.5\% & 55.3\% & 57.8\%\\
reply to someone & 62.7\% & 61.6\% & 70.3\% & 75.9\% & 82.4\%\\
components $c$ (final) & 2{,}871 & 1{,}036 & 722 & 135 & 214\\
cycle rank $\bone$ (final) & 35{,}936 & 11{,}882 & 55{,}087 & 40{,}525 & 228{,}772\\
cycle rate $\rho$ & 0.362 & 0.363 & 0.544 & 0.626 & 0.742\\
\bottomrule
\end{tabular}\end{adjustbox}
\end{table}

%\subsection{Instantiating the construction for Stack Exchange}
We instantiate the general construction of \cref{sec:online} as follows.
Vertices are users and hyperedges are threads; a thread consists of a question, its answers, and the comments on
the question and on the answers. A post by a user who is not yet a participant of the thread creates the event
$(u,t,a)$, where the anchor $a$ is the incidence of the author of the parent post (the question for an answer,
the commented post for a comment) in the same thread, and $a=\varnothing$ for the question itself. With these
anchors, the anchor rule of \cref{sec:rules} becomes the \emph{reply rule}: a new participant is placed next to
the person they reply to. Later posts by the same user in the same thread create no event, because the incidence already
exists. Posts by deleted accounts (without a user identifier) are dropped. Events are sorted by creation time.

%\subsection{Protocol}
For every community we run the growing dessin on the full history with the reply rule (main setting), the hub
rule and the random rule, two runs each with different random seeds (the seeds affect only tie-breaking), and
the shuffled ablation with two independent shufflings. We record the counts after every 1\% of the events. For
the bracketing experiment we draw five random samples of $1{,}000$ threads with at least two participants per
community and compute, on each sample, the growing dessin with the random rule (floor), with the reply rule
(main) and the offline face-maximization ceiling.

%\subsection{Reporting}
We report all measures as functions of calendar time and of $\bone$, and tabulate them at matched $\bone\in\{5{,}000,10{,}000,20{,}000\}$ (Artificial Intelligence ends at $\bone=11{,}882$ and has no value at $20{,}000$; Academia's first recorded point already has $\bone\approx2{,}300$, so we report no matched values below $5{,}000$). The full range of the final $\nloc$ across seeds is at most $0.3$ (main rule), as shown by shaded bands in the figures.

\section{Results}
\label{sec:results}

In this section, we apply the growing dessin to the five communities and read off how their re-meeting is organized. We first provide an overview of the main setting, the actual order of events under the reply rule, and compare the communities at matched numbers of cycles (\cref{sec:overview}). We then describe what the dessin reveals in each community (\cref{sec:per-community}). The remaining subsections test the method itself. A negative control shows that rotations read directly from the data behave like random maps, which motivates the growing construction (\cref{sec:negative}). We then check that the main conclusions are robust to the insertion rule (\cref{sec:robustness}), use a shuffled ablation to separate the effect of the real order of events from that of placement (\cref{sec:ablation}), and bracket the main rule between a random floor and an offline face-maximization ceiling (\cref{sec:bracket}).

\subsection{Overview}
\label{sec:overview}

\begin{figure}[!htbp]
\centering
\includegraphics[width=\linewidth]{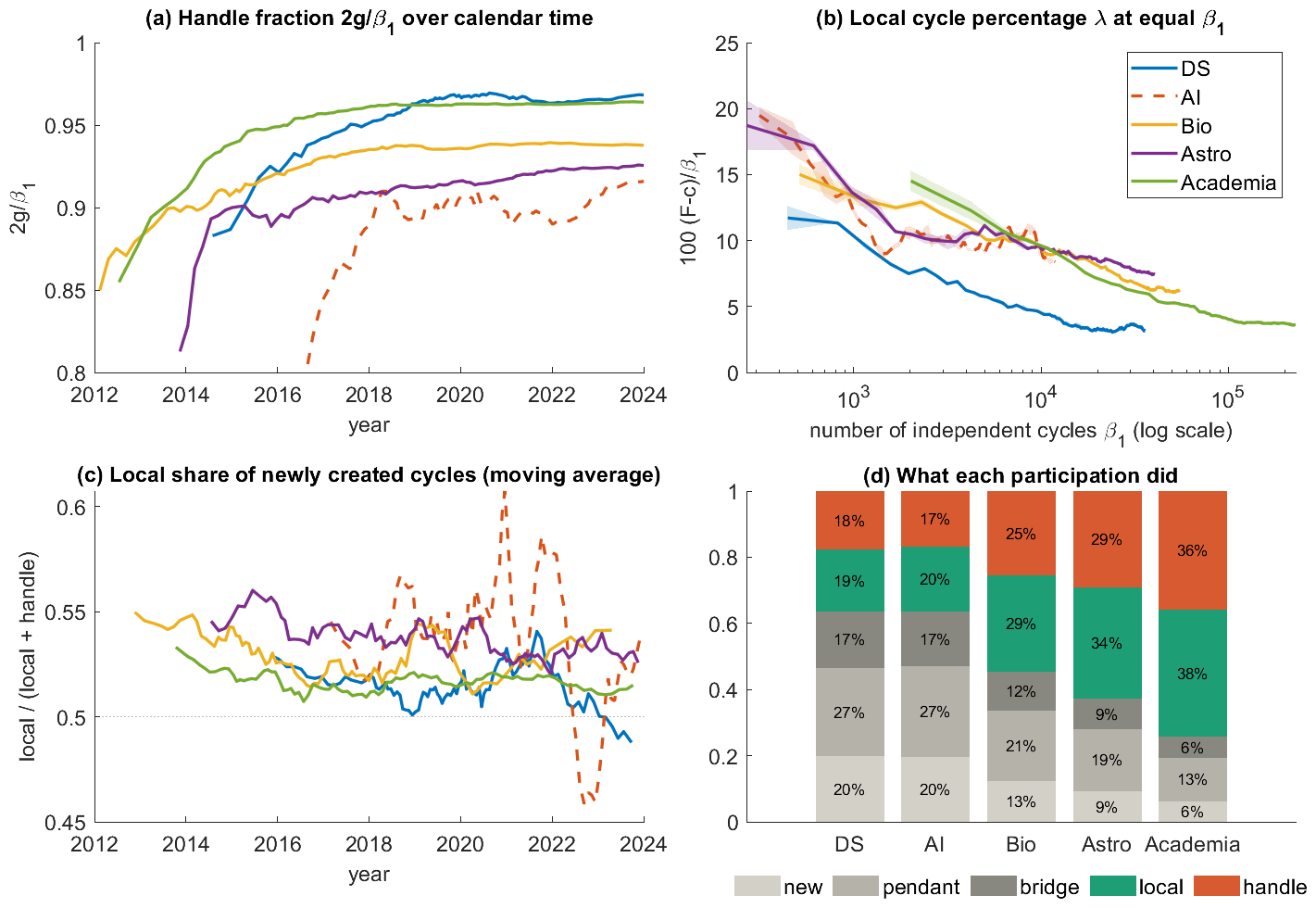}
\caption{Five communities under the main setting (real order, reply rule). (a) Handle fraction over calendar
time. (b) Local cycle percentage as a function of the number of cycles, the size-controlled comparison;
bands show the range over runs. (c) Local share of newly created cycles in successive windows (seven-point
moving average; $0.5$ means as many new handles as new local faces). (d) Composition of all events.}
\label{fig:main}
\end{figure}

\begin{table}[t]
\centering
\caption{Local cycle percentage, $\nloc=100(F-c)/\bone=100(1-\hfrac)$, at matched numbers of cycles and at the end of each history, for the three online rules and the shuffled ablation (means over runs). The date at which each community reached the given $\bone$ under the main rule is shown in the last block.}
\label{tab:matched}
\small
\begin{adjustbox}{max width=\linewidth}\begin{tabular}{@{}llrrrrr@{}}
\toprule
rule & $\bone$ & Data Science & Artif.\ Intell. & Biology & Astronomy & Academia\\
\midrule
\multirow{4}{*}{reply (main)} & 5{,}000 & {5.7} & 9.4 & 10.1 & 10.7 & 10.6\\
 & 10{,}000 & {4.5} & 9.2 & 9.2 & 9.3 & 9.4\\
 & 20{,}000 & {3.3} & -- & 8.1 & 8.4 & 7.0\\
 & end & {3.2} & 8.4 & 6.3 & 7.4 & 3.6\\
\midrule
\multirow{4}{*}{hub} & 5{,}000 & {5.3} & 9.4 & 10.5 & 9.8 & 11.0\\
 & 10{,}000 & {4.3} & 9.3 & 9.3 & 9.1 & 9.4\\
 & 20{,}000 & {3.1} & -- & 8.2 & 8.4 & 7.0\\
 & end & {3.2} & 8.4 & 6.2 & 7.4 & 3.5\\
\midrule
\multirow{4}{*}{random} & 5{,}000 & {4.7} & 8.4 & 8.9 & 9.8 & 10.0\\
 & 10{,}000 & {3.9} & 7.9 & 8.0 & 8.5 & 8.6\\
 & 20{,}000 & {2.7} & -- & 6.6 & 7.4 & 6.2\\
 & end & {2.7} & 7.0 & 5.2 & 6.6 & 3.2\\
\midrule
\multirow{4}{*}{shuffled + random} & 5{,}000 & 2.4 & 5.4 & 4.5 & 5.9 & 2.7\\
 & 10{,}000 & 2.3 & 5.0 & 4.6 & 6.1 & 2.6\\
 & 20{,}000 & 2.2 & -- & 4.6 & 6.2 & 2.7\\
 & end & 2.0 & 5.1 & 4.8 & 7.2 & 4.5\\
\midrule
date reached & 10{,}000 & 2018-05 & 2023-02 & 2014-08 & 2017-03 & 2013-10\\
(main rule) & 20{,}000 & 2020-01 & -- & 2015-11 & 2019-11 & 2014-07\\
\bottomrule
\end{tabular}\end{adjustbox}
\end{table}

\begin{table}[t]
\centering
\caption{Final state of each community under the main rule (means over two runs).}
\label{tab:final}
\small
\begin{adjustbox}{max width=\linewidth}\begin{tabular}{@{}lrrrrr@{}}
\toprule
 & Data Science & Artif.\ Intell. & Biology & Astronomy & Academia\\
\midrule
cycle rank $\bone$ & 35{,}936 & 11{,}882 & 55{,}087 & 40{,}525 & 228{,}772\\
genus $g$ ($=H$) & 17{,}401 & 5{,}443 & 25{,}820 & 18{,}757 & 110{,}259\\
local events $L$ & 18{,}536 & 6{,}439 & 29{,}267 & 21{,}768 & 118{,}514\\
faces $F$ & 4{,}006 & 2{,}032 & 4{,}169 & 3{,}146 & 8{,}469\\
components $c$ & 2{,}871 & 1{,}036 & 722 & 135 & 214\\
handle fraction $\hfrac$ & 0.968 & 0.916 & 0.937 & 0.926 & 0.964\\
$\nloc$ & 3.2 & 8.4 & 6.3 & 7.4 & 3.6\\
\midrule
new / pendant / bridge (\%) & 20 / 27 / 17 & 20 / 27 / 17 & 13 / 21 / 12 & 9 / 19 / 9 & 6 / 13 / 6\\
local / handle (\%) & 19 / 18 & 20 / 17 & 29 / 25 & 34 / 29 & 38 / 36\\
\bottomrule
\end{tabular}\end{adjustbox}
\end{table}

\Cref{fig:main} and \cref{tab:matched} summarize the main setting. Four observations frame the community-level results. First, all communities are close to the random-map regime but measurably away from it: final handle fractions lie between $0.92$ and $0.97$, whereas a random map has a handle fraction close to $1$. Because these values are so close to one another, we instead report the distance from $1$, rescaled as $\nloc=100(1-\hfrac)$. Second, local structure declines with size in every community (\cref{fig:main}b), so raw final values confound size and structure; Academia's final handle fraction ($0.964$) is close to Data Science's ($0.968$) only because it is six times larger in $\bone$. Third, after an initial rise, the handle fraction reaches a plateau in every community except Astronomy, which rises slowly throughout (\cref{fig:main}a). Fourth, every community closes more cycles locally than globally, as \cref{cor:conservation} requires, but only slightly: the local share of new cycles stays between about $0.50$ and $0.56$ in most windows (\cref{fig:main}c).

\subsection{What the dessin reveals in each community}
\label{sec:per-community}

\paragraph{Data Science: tangled growth.}
Data Science is the most diffuse community: threads are small ($2.70$ participants), only $6.4\%$ of co-participating pairs ever meet again, and $64\%$ of participations close no cycle (\cref{tab:final}). The dessin adds a statement about \emph{how} people meet when they do: at every matched size and under every rule, Data Science closes about half as many cycles locally as any other community ($\nloc=4.5$ against $9.2$--$9.4$ at $\bone=10{,}000$; \cref{tab:matched}), so its re-meetings usually join parts of the community that were not adjacent on the surface. Its handle fraction rises until 2019 and then stays near $0.97$ (\cref{fig:main}a). In the last two years, the local share of new cycles even dips below $0.5$ in some windows (\cref{fig:main}c), which is possible in a window although never cumulatively (\cref{cor:conservation}); these windows, however, contain fewer events as activity declines.

\paragraph{Artificial Intelligence: same statistics, different surface.}
Artificial Intelligence is the most informative comparison. Its order-free profile is almost identical to Data Science's (\cref{tab:case}): participants per thread, the share of threads with at least three participants, the share of one-time users, the cycle rate and the event composition all agree. Yet at equal numbers of cycles, it closes twice as many locally ($9.2$ versus $4.5$ at $\bone=10{,}000$), under every rule, including the shuffled ablation ($5.0$ versus $2.3$). Only the repeat-pair rate hints at the difference ($9.2\%$ versus $6.4\%$), and the dessin shows what it means: in Artificial Intelligence, re-meetings happen among people who are already neighbors on the surface, while in Data Science the same amount of re-meeting is spread across the community. Its handle fraction, noisy because it is the smallest community, rises from 2023 onward (\cref{fig:main}a), as an influx of newcomers connecting distant regions would produce.

\begin{table}[t]
\centering
\caption{Case study: Data Science versus Artificial Intelligence. Standard statistics are nearly identical;
the dessin measure differs by a factor of two.}
\label{tab:case}
\small
\begin{adjustbox}{max width=\linewidth}\begin{tabular}{@{}lrr@{}}
\toprule
 & Data Science & Artificial Intelligence\\
\midrule
participants per thread & 2.70 & 2.65\\
threads with $\ge3$ participants & 43.8\% & 43.8\%\\
users in a single thread & 69.5\% & 69.5\%\\
cycle rate $\rho$ & 0.362 & 0.363\\
events new / pendant / bridge / local / handle (\%) & 20 / 27 / 17 / 19 / 18 & 20 / 27 / 17 / 20 / 17\\
repeat pairs & 6.4\% & 9.2\%\\
\midrule
$\nloc$ at $\bone=5{,}000$ (main rule) & \textbf{5.7} & \textbf{9.4}\\
$\nloc$ at $\bone=10{,}000$ (main rule) & \textbf{4.5} & \textbf{9.2}\\
$\nloc$ at $\bone=10{,}000$ (shuffled ablation) & 2.3 & 5.0\\
\bottomrule
\end{tabular}\end{adjustbox}
\end{table}

\paragraph{Biology: a mature community with a stable core.}
Biology has larger threads ($3.52$ participants), a strong core ($54.5\%$ of participations by the top 1\% of users) and a handle fraction essentially flat at $0.937$ since 2018: a steady state in which about one cycle in sixteen is absorbed as a local face beyond the random-map level. It gains, with Artificial Intelligence, the most from the reply rule ($+1.2$ at $\bone=10{,}000$), so its reply structure predicts where future cycles close, and it depends strongly on the real temporal order ($+3.4$).

\paragraph{Astronomy: dense threads that reconnect regardless of timing.}
Astronomy has dense threads ($4.43$ participants) and the most connected hypergraph ($135$ components at the end), and it keeps the most local structure at large sizes ($8.4$ at $\bone=20{,}000$). Its local structure depends least on timing: the shuffled ablation retains $6.1$ local cycle percentage at $\bone=10{,}000$, the highest of all, consistent with large threads that repeatedly bring the same people together whatever the order.

\paragraph{Academia: discussion-heavy, highly sequential.}
Academia's threads are long discussions ($7.04$ participants, $89\%$ with at least three), and $74\%$ of its participations close a cycle. Its local structure is among the highest at small sizes ($10.6$ at $\bone=5{,}000$) but declines fastest with growth ($7.0$ at $20{,}000$ and $3.6$ at the end). Most strikingly, it depends most on the real order of events: at $\bone=10{,}000$, replaying the same interactions in shuffled order leaves only $2.6$ local cycle percentage, against $8.6$ in the real order with the same random rule. In a discussion community, \emph{when} people come back matters.
%=====================================================================================================
\subsection{Negative control: rotations read from the data}
\label{sec:negative}
 
We now test the claim that rotations read directly from the data behave like random maps. The static construction orders the participants of a thread by their first post (or by the reply tree, through a depth-first walk around question $\to$ answers $\to$ comments) and the threads of a user by the time the user joined them.

\begin{figure}[!htbp]
\centering
\includegraphics[width=\linewidth]{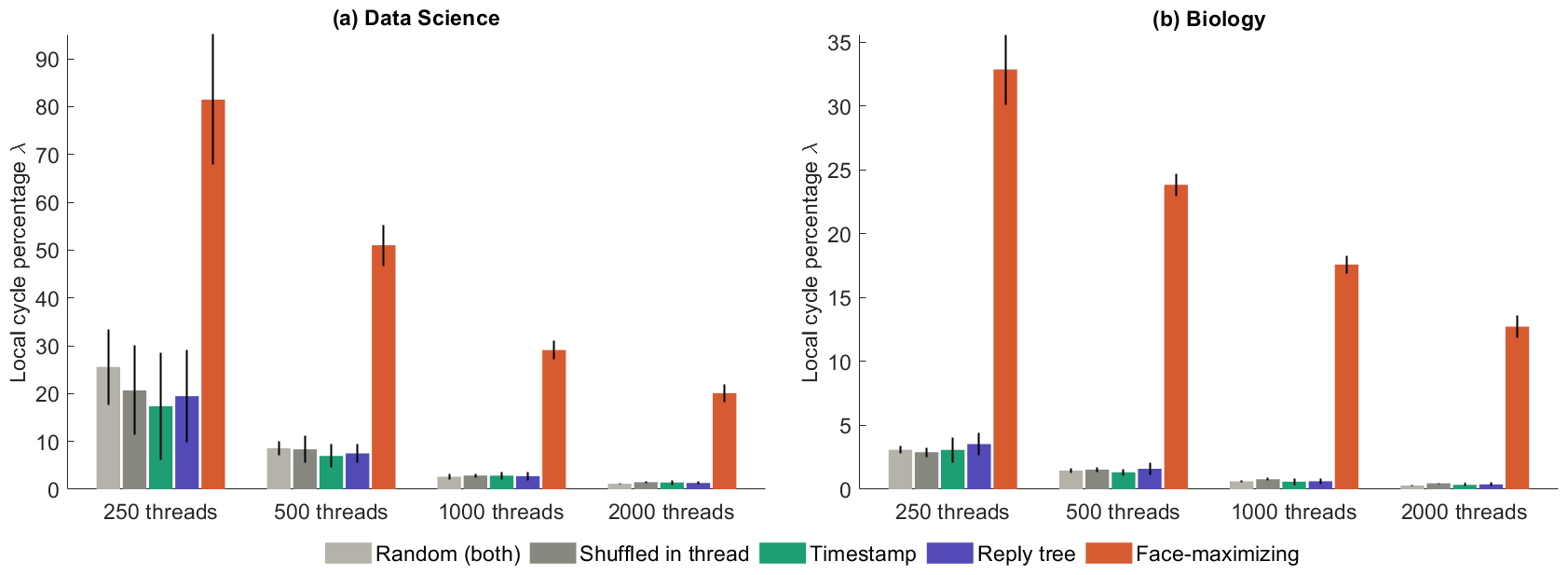}
\caption{Negative control: rotations read from the data, for (a) Data Science and (b) Biology. For random samples of 250--2{,}000 threads, each sample is a fixed hypergraph, and we compare five rotation systems on it by the share of its independent cycles that are local, measured as local cycle percentage, $\nloc=100\,(F-c)/\bone$ (\cref{sec:measures}); bars show the mean and error bars the standard deviation over five samples. Timestamp and reply-tree orders are indistinguishable from random orders; only face maximization produces substantial local structure.}
\label{fig:negative}
\end{figure}

We drew random samples of $250$, $500$, $1{,}000$ and $2{,}000$ threads from Data Science and Biology, kept the
hypergraph of each sample fixed, and compared five rotation systems: random rotations at both users and
threads; time order at users and random order inside threads; time order at both; time order at users and
reply-tree order inside threads; and the offline face-maximizing rotation system. \Cref{fig:negative} shows the
local cycle percentage. For the largest samples, Data Science ($\bone\approx860$) gives $1.1$ (random),
$1.5$ (shuffled inside threads), $1.4$ (timestamps), $1.3$ (reply tree) and $20.1$ (face maximization); Biology
($\bone\approx2{,}780$) gives $0.3$, $0.4$, $0.4$, $0.4$ and $12.7$. Paired tests show no meaningful
differences between timestamp, reply-tree, and shuffled orders. Moreover, the reply-tree order differs from
the timestamp order (as a cyclic order) in only $7.6\%$ of Data Science threads with at least three
participants and $12.9\%$ of Biology threads.
 
Used as cyclic orders, timestamps, and reply trees thus produce no more faces than random orders do, although the face-maximizing rotation system shows that the same hypergraphs admit far more local structure.

\subsection{Robustness to the insertion rule}
\label{sec:robustness}

\begin{figure}[!htbp]
\centering
\includegraphics[width=\linewidth]{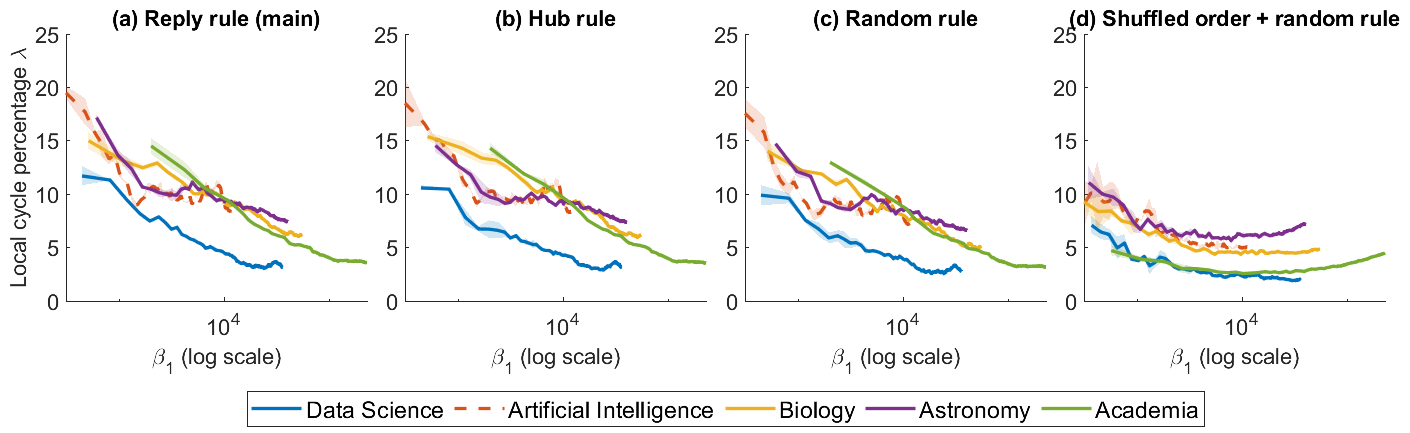}
\caption{Local cycle percentage, $\nloc=100\,(F-c)/\bone$, during the growth of each community's dessin, for the three online insertion rules and the shuffled ablation. Each curve follows one community through its history: a point at a given value of $\bone$ shows the share of all cycles created up to that moment that are local, so communities are compared at equal size rather than at equal calendar time. Lines show the mean over runs with different random seeds, and bands the range. Data Science is the lowest community under every rule.}
\label{fig:rules}
\end{figure}

\Cref{fig:rules} and \cref{tab:matched} compare the three online rules, and the qualitative picture is the same under all of them: Data Science is lowest by a factor of about two at every matched size. The other four communities lie within about $1$ point of each other, and their order changes between rules and sizes, so we make no claim about their ranking. The reply and hub rules give nearly the same values, and the random rule lowers every community by about $0.6$--$1.2$ points at $\bone=10{,}000$. The statement that Data Science grows in a more tangled way than the other four communities is robust to the insertion rule; finer rankings are not.

\subsection{Ablation: does the real order of events matter?}
\label{sec:ablation}

\begin{figure}[!htbp]
\centering
\includegraphics[width=\linewidth]{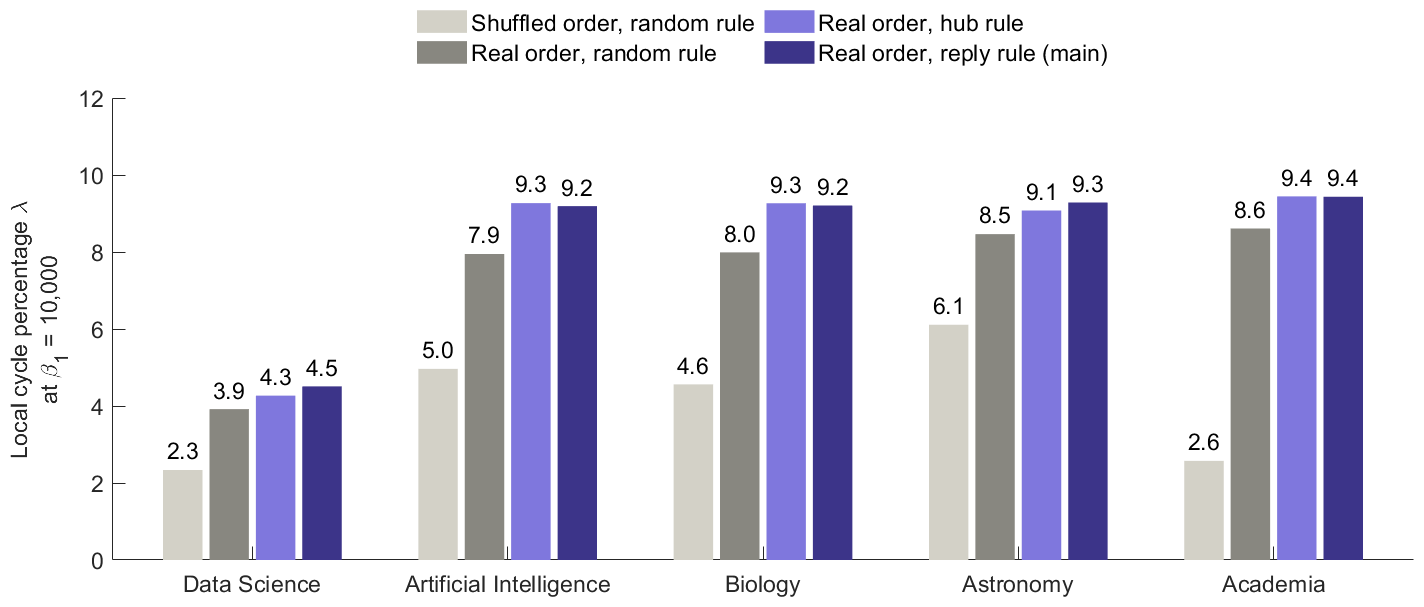}
\caption{Local cycle percentage, $\nloc=100\,(F-c)/\bone$, read off each community's growing dessin at the moment it reaches $\bone=10{,}000$ cycles (a vertical slice through \cref{fig:rules}), averaged over runs. From left to right within each community: shuffled order with the random rule; real order with the random rule; real order with the hub rule; real order with the reply rule (main). The first two bars isolate the effect of the real time order, and the last three the effect of the insertion rule.}
\label{fig:ablation}
\end{figure}

\Cref{fig:ablation} separates the contributions of time order and of placement at $\bone=10{,}000$. Comparing the real order with a shuffled order of the same events (both generated by the same random rule) isolates the effect of history. Real order adds $+1.6$ local cycle percentage in Data Science, $+3.0$ in Artificial Intelligence, $+3.4$ in Biology, $+2.4$ in Astronomy and $+6.0$ in Academia (differences are computed before rounding). In every community, the actual sequence of interactions builds more local structure than a random sequence would, and the effect is largest in the discussion community.

Comparing the reply rule with the random rule on the real order isolates the effect of where newcomers are placed: $+0.6$ (Data Science), $+1.2$ (Artificial Intelligence), $+1.2$ (Biology), $+0.8$ (Astronomy) and $+0.8$ (Academia). Placing newcomers next to the person they reply to consistently helps future cycles close locally, although the effect is smaller than the effect of time order.

At the \emph{end} of their histories, the comparison is not uniform (\cref{tab:matched}, rows ``end''): with the random rule, the final $\nloc$ is higher for the real order in Data Science, Artificial Intelligence and Biology, but higher for the shuffled order in Astronomy ($7.2$ versus $6.6$) and Academia ($4.5$ versus $3.2$). A plausible explanation, which we leave for future work, is that real histories close many local loops early, when a small core of regulars interacts repeatedly, while later growth onto a large established surface produces mostly handles; a shuffled history spreads the core's interactions evenly over time.

\subsection{Bracketing: floor, main rule and ceiling}
\label{sec:bracket}

\begin{figure}[!htbp]
\centering
\includegraphics[width=\linewidth]{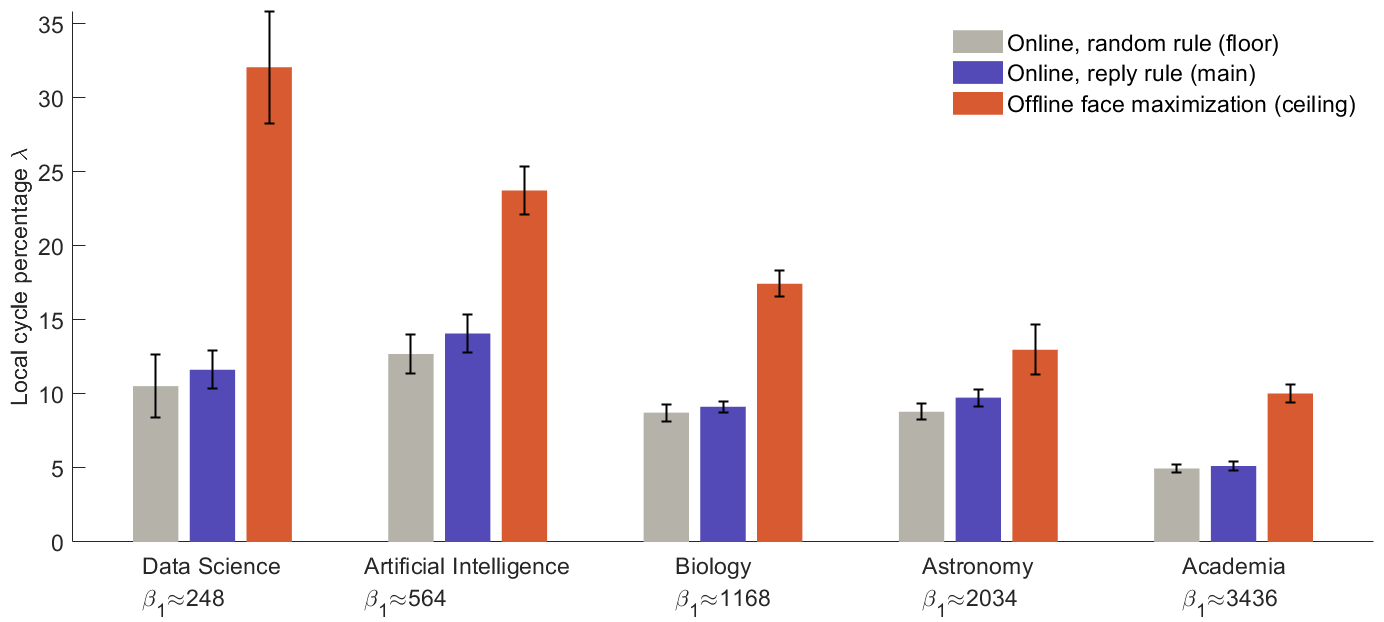}
\caption{Floor, main setting, and ceiling on the same hypergraphs. From each community we draw five random samples of 1{,}000 threads and compute the local cycle percentage, $\nloc=100\,(F-c)/\bone$, of three dessins of each sample: the dessin grown online in the real order with the random rule (floor), the same with the reply rule (main), and an offline face-maximizing rotation system of the final sample hypergraph, found by search without regard to time (ceiling; a lower bound on the true maximum). Bars show the mean, and error bars indicate the standard deviation across the five samples. Because samples of 1{,}000 threads have very different numbers of cycles in different communities ($\bone$ is shown under each group) and $\nloc$ decreases with size, bars should be compared \emph{within} a community.}
\label{fig:bracket}
\end{figure}

\Cref{fig:bracket} places the online results between a floor and a ceiling computed on the same sub-hypergraphs.
In every community, the reply rule is at or above the random rule ($11.6$ versus $10.5$ in Data Science, $14.1$ versus $12.7$ in Artificial Intelligence, $9.1$ versus $8.7$ in Biology, $9.7$ versus $8.8$ in Astronomy, $5.1$ versus $5.0$ in Academia), although at this small scale the differences are within about one standard deviation. The offline ceiling is $1.3$ to $2.8$ times higher ($32.1$, $23.7$, $17.4$, $13.0$ and $10.0$), so the growing dessin realizes roughly $35\%$ (Data Science) to $75\%$ (Astronomy) of the local structure that could be closed in hindsight. This suggests, as a hypothesis given that the ceiling is a heuristic lower bound, that the communities differ less in which cycles \emph{could} be local than in how their growth realizes them.

\subsection{Summary of findings}

Under every online insertion rule, Data Science organizes its interaction cycles in a far more tangled way than the other four communities, with about half their local structure at equal size. This difference is invisible to standard statistics: Data Science and Artificial Intelligence are indistinguishable in participants per thread, single-thread users and cycle rate, yet they differ by a factor of two in the dessin measure. The temporal order of the events matters as well. At matched size, the real order creates more local structure than shuffled histories of the same hypergraph, by $1.6$ to $6.0$ points, most strongly in the discussion community, Academia; placing newcomers by reply structure adds a further $0.6$ to $1.2$ points. Finally, online growth realizes 35--75\% of the local structure that is available in hindsight, as measured by the offline face-maximization ceiling.

%=====================================================================================================
\section{Discussion}
\label{sec:discussion}

\subsection{What dessins bring to hypergraph analysis}
Beyond the results, the dessin viewpoint adds three things to hypergraph analysis. Its faces are a third kind of cell that is \emph{forced} by the data rather than chosen by a filling rule, and triality makes them a natural partner of vertices and hyperedges. It decomposes re-meeting exactly, $\bone=(F-c)+2g$, so that communities with the same amount of re-meeting can differ in how it is organized, and the event calculus with its conservation law turns this decomposition into time-resolved trajectories rather than unrelated snapshots. Finally, because the same final hypergraph can grow into different surfaces, comparing real and shuffled histories quantifies the topological role of temporal order, which order-free statistics cannot do.

\subsection{Limitations}
The main limitation is that the surface depends on where darts are inserted. We fix the rule in advance, test three online rules and a shuffled ablation, and bracket the results between a random floor and an offline ceiling; the main conclusions hold under all of them, but absolute values and finer rankings do not. The genus of a growing dessin is that of a particular embedding; hence, there is an upper bound on the minimum genus, and the shared-face search is only myopically optimal, as the toy example shows. The ceiling is heuristic, because face maximization is NP-hard~\citep{Thomassen1989}, and depends on the search budget for large samples. With five communities, the comparison is descriptive; statistical statements need a larger study across many sites and domains. Finally, modeling choices (first participations only, comments as participation, dropped deleted accounts) affect the hypergraph.

\subsection{Future directions}
\label{sec:future}
The most direct next step is a dessin neural network: darts carry natural features (a post's time, role, and content), the operators $\sigma_0$, $\sigma_1$, and $\sinf$ define a message-passing scheme that is equivariant under dart relabeling, the growing dessin provides the rotations for temporal data, and faces provide a third pooling level. Since the event calculus is defined per event, each vertex also has its own trajectory of local and handle events, making it a natural route to questions such as early detection of future experts. The construction can be sharpened by learning the insertion rule and computing the exact minimum genera of small hypergraphs using integer programming. On the mathematical side, coverings of dessins give principled hierarchical pooling with controlled changes in genus via Riemann--Hurwitz, and Galois-conjugate dessins are natural hard instances for testing the expressive power of hypergraph learning models. Finally, applying the construction to the domains of \cref{tab:domains} would show which of our findings are specific to question--answer communities.
%=====================================================================================================
\section{Conclusion}
\label{sec:conclusion}
A hypergraph with ordered incidences is a dessin d'enfant, and therefore a surface. We used this classical correspondence to define the growing dessin, a topological description of any temporal hypergraph in which every new incidence is inserted into a surface built from the history itself, and an exact event classification with a conservation law makes it interpretable: every new cycle closes either locally, as a new face, or globally, as a new handle. In a first application to five Stack Exchange communities, this description separates communities that standard statistics cannot distinguish, quantifies how much the real temporal order of interactions matters, and is robust to the choice of insertion rule in its main conclusions. We hope that this first step will bring the mathematics of dessins d'enfants, with its Galois symmetries, coverings and Belyi maps, into the study of higher-order networks and their learning.

\section*{Data and code availability}
All data are from the public Stack Exchange data dump~\citep{StackExchangeDump} (CC BY-SA 4.0). The implementation (growing dessin, insertion rules, offline face maximization, and all experiment scripts) is available at \url{https://github.com/mehmeteaktas/dessins-temporal-hypergraphs}.
\section*{Acknowledgments}
This work was supported in part by the National Science Foundation under Grant No.~2523660. The authors used Claude (Anthropic), an AI assistant, during the preparation of this paper, for help with code for the experiments, figures, and the drafting and editing of text. The authors reviewed, verified, and edited all content and take full responsibility for the paper.

\bibliographystyle{plainnat}
\bibliography{refs}

@incollection{Grothendieck1997Esquisse,
  author    = {Grothendieck, Alexander},
  title     = {Esquisse d'un programme},
  booktitle = {Geometric Galois Actions, Vol. 1},
  editor    = {Schneps, Leila and Lochak, Pierre},
  series    = {London Mathematical Society Lecture Note Series},
  volume    = {242},
  publisher = {Cambridge University Press},
  pages     = {5--48},
  year      = {1997},
  note      = {Written in 1984}
}

@article{aktas2026persistent,
  title={Persistent homology of hypergraphs via barycentric subdivision},
  author={Aktas, Mehmet Emin and Nguyen, Thu and Riza, Rakin and Islam, Muhammad Ifte and Guler, Ahmed and Akbas, Esra},
  journal={Journal of Complex Networks},
  volume={14},
  number={3},
  pages={cnag013},
  year={2026},
  publisher={Oxford University Press}
}

@article{gasparovic2026survey,
  title={A survey of simplicial, relative, and chain complex homology theories for hypergraphs},
  author={Gasparovic, Ellen and Purvine, Emilie and Sazdanovic, Radmila and Wang, Bei and Wang, Yusu and Ziegelmeier, Lori},
  journal={Journal of Applied and Computational Topology},
  volume={10},
  number={2},
  pages={7},
  year={2026},
  publisher={Springer}
}

@article{Belyi1979,
  author  = {Bely{\u\i}, G. V.},
  title   = {On {G}alois extensions of a maximal cyclotomic field},
  journal = {Izvestiya Akademii Nauk SSSR. Seriya Matematicheskaya},
  volume  = {43},
  number  = {2},
  pages   = {267--276},
  year    = {1979}
}

@article{aktas2017dessins,
  title={Dessins d'Enfants of Trigonal Curves},
  author={Aktas, Mehmet Emin},
  journal={arXiv preprint arXiv:1706.09956},
  year={2017}
}

@inproceedings{aktas2025expert,
  title={Expert detection on crowdsourcing forums via sheaf {L}aplacian},
  author={Aktas, Mehmet Emin and Moradi, Iraj and Cosar, Ibrahim and Goyal, Yateeka and Alyaz, Asya and Akbas, Esra and Ahsen, Mehmet},
  booktitle={2025 IEEE 12th International Conference on Data Science and Advanced Analytics (DSAA)},
  pages={1--10},
  year={2025},
  organization={IEEE}
}

@inproceedings{kusakula2025beyond,
  title={Beyond Pairwise Links: Hypergraph Modeling for Scientific Trend Forecasting},
  author={Kusakula, Nithyasree and Gangavaram, Navyamsh and Torres, Liz and Funk, Russell J and Aktas, Mehmet Emin},
  booktitle={International Conference on Advances in Social Networks Analysis and Mining},
  pages={298--312},
  year={2025},
  organization={Springer}
}

@book{LandoZvonkin2004,
  author    = {Lando, Sergei K. and Zvonkin, Alexander K.},
  title     = {Graphs on Surfaces and Their Applications},
  series    = {Encyclopaedia of Mathematical Sciences},
  volume    = {141},
  publisher = {Springer},
  year      = {2004}
}

@book{GirondoGonzalezDiez2012,
  author    = {Girondo, Ernesto and Gonz{\'a}lez-Diez, Gabino},
  title     = {Introduction to Compact {R}iemann Surfaces and Dessins d'Enfants},
  series    = {London Mathematical Society Student Texts},
  volume    = {79},
  publisher = {Cambridge University Press},
  year      = {2012}
}

@book{JonesWolfart2016,
  author    = {Jones, Gareth A. and Wolfart, J{\"u}rgen},
  title     = {Dessins d'Enfants on {R}iemann Surfaces},
  series    = {Springer Monographs in Mathematics},
  publisher = {Springer},
  year      = {2016}
}

@book{Schneps1994,
  editor    = {Schneps, Leila},
  title     = {The {G}rothendieck Theory of Dessins d'Enfants},
  series    = {London Mathematical Society Lecture Note Series},
  volume    = {200},
  publisher = {Cambridge University Press},
  year      = {1994}
}

@article{Walsh1975,
  author  = {Walsh, T. R. S.},
  title   = {Hypermaps versus bipartite maps},
  journal = {Journal of Combinatorial Theory, Series B},
  volume  = {18},
  number  = {2},
  pages   = {155--163},
  year    = {1975}
}

@article{Cori1975,
  author  = {Cori, Robert},
  title   = {Un code pour les graphes planaires et ses applications},
  journal = {Ast{\'e}risque},
  volume  = {27},
  year    = {1975}
}

@article{JonesSingerman1978,
  author  = {Jones, Gareth A. and Singerman, David},
  title   = {Theory of maps on orientable surfaces},
  journal = {Proceedings of the London Mathematical Society},
  volume  = {s3-37},
  number  = {2},
  pages   = {273--307},
  year    = {1978}
}

@book{MoharThomassen2001,
  author    = {Mohar, Bojan and Thomassen, Carsten},
  title     = {Graphs on Surfaces},
  publisher = {Johns Hopkins University Press},
  year      = {2001}
}

@book{GrossTucker1987,
  author    = {Gross, Jonathan L. and Tucker, Thomas W.},
  title     = {Topological Graph Theory},
  publisher = {Wiley},
  year      = {1987}
}

@article{Edmonds1960,
  author  = {Edmonds, Jack},
  title   = {A combinatorial representation for polyhedral surfaces},
  journal = {Notices of the American Mathematical Society},
  volume  = {7},
  pages   = {646},
  year    = {1960}
}

@article{Thomassen1989,
  author  = {Thomassen, Carsten},
  title   = {The graph genus problem is {NP}-complete},
  journal = {Journal of Algorithms},
  volume  = {10},
  number  = {4},
  pages   = {568--576},
  year    = {1989}
}

@article{Xuong1979,
  author  = {Xuong, Nguyen Huy},
  title   = {How to determine the maximum genus of a graph},
  journal = {Journal of Combinatorial Theory, Series B},
  volume  = {26},
  number  = {2},
  pages   = {217--225},
  year    = {1979}
}

@article{HarerZagier1986,
  author  = {Harer, John and Zagier, Don},
  title   = {The {E}uler characteristic of the moduli space of curves},
  journal = {Inventiones Mathematicae},
  volume  = {85},
  number  = {3},
  pages   = {457--485},
  year    = {1986}
}

@article{PippengerSchleich2006,
  author  = {Pippenger, Nicholas and Schleich, Kristin},
  title   = {Topological characteristics of random triangulated surfaces},
  journal = {Random Structures \& Algorithms},
  volume  = {28},
  number  = {3},
  pages   = {247--288},
  year    = {2006}
}

@article{Gamburd2006,
  author  = {Gamburd, Alex},
  title   = {Poisson--{D}irichlet distribution for random {B}elyi surfaces},
  journal = {The Annals of Probability},
  volume  = {34},
  number  = {5},
  pages   = {1827--1848},
  year    = {2006}
}

@article{Dixon1969,
  author  = {Dixon, John D.},
  title   = {The probability of generating the symmetric group},
  journal = {Mathematische Zeitschrift},
  volume  = {110},
  number  = {3},
  pages   = {199--205},
  year    = {1969}
}

@article{Battiston2020,
  author  = {Battiston, Federico and Cencetti, Giulia and Iacopini, Iacopo and Latora, Vito and Lucas, Maxime and Patania, Alice and Young, Jean-Gabriel and Petri, Giovanni},
  title   = {Networks beyond pairwise interactions: Structure and dynamics},
  journal = {Physics Reports},
  volume  = {874},
  pages   = {1--92},
  year    = {2020}
}

@article{Bick2023,
  author  = {Bick, Christian and Gross, Elizabeth and Harrington, Heather A. and Schaub, Michael T.},
  title   = {What are higher-order networks?},
  journal = {SIAM Review},
  volume  = {65},
  number  = {3},
  pages   = {686--731},
  year    = {2023}
}

@book{Berge1973,
  author    = {Berge, Claude},
  title     = {Graphs and Hypergraphs},
  publisher = {North-Holland},
  year      = {1973}
}

@article{Benson2018,
  author  = {Benson, Austin R. and Abebe, Rediet and Schaub, Michael T. and Jadbabaie, Ali and Kleinberg, Jon},
  title   = {Simplicial closure and higher-order link prediction},
  journal = {Proceedings of the National Academy of Sciences},
  volume  = {115},
  number  = {48},
  pages   = {E11221--E11230},
  year    = {2018}
}

@inproceedings{Paranjape2017,
  author    = {Paranjape, Ashwin and Benson, Austin R. and Leskovec, Jure},
  title     = {Motifs in temporal networks},
  booktitle = {Proceedings of the Tenth ACM International Conference on Web Search and Data Mining (WSDM)},
  pages     = {601--610},
  year      = {2017}
}

@article{Carlsson2009,
  author  = {Carlsson, Gunnar},
  title   = {Topology and data},
  journal = {Bulletin of the American Mathematical Society},
  volume  = {46},
  number  = {2},
  pages   = {255--308},
  year    = {2009}
}

@book{EdelsbrunnerHarer2010,
  author    = {Edelsbrunner, Herbert and Harer, John},
  title     = {Computational Topology: An Introduction},
  publisher = {American Mathematical Society},
  year      = {2010}
}

@article{Aktas2019,
  author  = {Aktas, Mehmet E. and Akbas, Esra and El Fatmaoui, Ahmed},
  title   = {Persistence homology of networks: methods and applications},
  journal = {Applied Network Science},
  volume  = {4},
  number  = {1},
  pages   = {61},
  year    = {2019}
}

@article{HansenGhrist2019,
  author  = {Hansen, Jakob and Ghrist, Robert},
  title   = {Toward a spectral theory of cellular sheaves},
  journal = {Journal of Applied and Computational Topology},
  volume  = {3},
  number  = {4},
  pages   = {315--358},
  year    = {2019}
}

@inproceedings{Bodnar2022,
  author    = {Bodnar, Cristian and Di Giovanni, Francesco and Chamberlain, Benjamin Paul and Li{\`o}, Pietro and Bronstein, Michael M.},
  title     = {Neural Sheaf Diffusion: A Topological Perspective on Heterophily and Oversmoothing in {GNN}s},
  booktitle = {Advances in Neural Information Processing Systems},
  year      = {2022}
}

@article{Hajij2022,
  author  = {Hajij, Mustafa and Zamzmi, Ghada and Papamarkou, Theodore and Miolane, Nina and Guzm{\'a}n-S{\'a}enz, Aldo and Ramamurthy, Karthikeyan Natesan and others},
  title   = {Topological Deep Learning: Going Beyond Graph Data},
  journal = {arXiv preprint arXiv:2206.00606},
  year    = {2022}
}

@inproceedings{Feng2019,
  author    = {Feng, Yifan and You, Haoxuan and Zhang, Zizhao and Ji, Rongrong and Gao, Yue},
  title     = {Hypergraph Neural Networks},
  booktitle = {Proceedings of the AAAI Conference on Artificial Intelligence},
  volume    = {33},
  pages     = {3558--3565},
  year      = {2019}
}

@inproceedings{ChitraRaphael2019,
  author    = {Chitra, Uthsav and Raphael, Benjamin J.},
  title     = {Random Walks on Hypergraphs with Edge-Dependent Vertex Weights},
  booktitle = {Proceedings of the 36th International Conference on Machine Learning (ICML)},
  pages     = {1172--1181},
  year      = {2019}
}

@inproceedings{Wang2023EDHNN,
  author    = {Wang, Peihao and Yang, Shenghao and Liu, Yunyu and Wang, Zhangyang and Li, Pan},
  title     = {Equivariant Hypergraph Diffusion Neural Operators},
  booktitle = {International Conference on Learning Representations (ICLR)},
  year      = {2023}
}

@misc{StackExchangeDump,
  author       = {{Stack Exchange, Inc.}},
  title        = {{S}tack {E}xchange Data Dump},
  howpublished = {Internet Archive, \url{https://archive.org/details/stackexchange}},
  note         = {Licensed under CC BY-SA 4.0},
  year         = {2024}
}

@book{Hatcher2002,
  author    = {Hatcher, Allen},
  title     = {Algebraic Topology},
  publisher = {Cambridge University Press},
  year      = {2002}
}

@article{Kirkpatrick1983,
  author  = {Kirkpatrick, Scott and Gelatt, C. Daniel and Vecchi, Mario P.},
  title   = {Optimization by simulated annealing},
  journal = {Science},
  volume  = {220},
  number  = {4598},
  pages   = {671--680},
  year    = {1983}
}

@article{Torres2021,
  author  = {Torres, Leo and Blevins, Ann S. and Bassett, Danielle and Eliassi-Rad, Tina},
  title   = {The why, how, and when of representations for complex systems},
  journal = {SIAM Review},
  volume  = {63},
  number  = {3},
  pages   = {435--485},
  year    = {2021}
}

\appendix
\section{Implementation details}
\label{sec:sm-impl}
\label{app:impl}

\subsection{Pseudocode of the lockstep retracing}
\label{sec:sm-lockstep}
Let $A$ be the set of at most four darts at which $\varphi$ changed, and let $O$ be the set of old face labels
of $A$. Start one tracer at each dart of $A$. In rounds, advance every active tracer by one application of
$\varphi$. When a tracer returns to its start, its cycle is complete: give it a new label and deactivate every
tracer whose start lies on it. When a tracer reaches the start of another active tracer, the two lie on the same
cycle and one of them is deactivated. Stop when at most one tracer is active. All completed cycles have been
relabeled; the remaining darts of the old faces in $O$ form a single cycle, which keeps the label of the largest
old face, and the other old faces in $O$ (if any) are merged into it. The total work is proportional to the
number of tracers times the length of the shortest completed cycle, plus the size of the merged faces that are
smaller than the largest one.

\subsection{Face maximization moves}
\label{sec:sm-moves}
A move at a thread takes a dart $x$ with predecessor $p=\sigma_1^{-1}(x)$ and reinserts it after another dart $y$
of the same thread. Then $\sigma_1'=\sigma_1\circ(p\;x\;y)$ and $\varphi'=\varphi\circ(p\;x\;y)$. Writing
$(p\;x\;y)=(p\;y)(p\;x)$ and applying \cref{lem:transposition} twice gives $\Delta F\in\{-2,0,+2\}$. Precisely: if $p$ and $y$ lie on the same face, then $\Delta F=+2$ when $x$
lies on that face strictly after $y$ and at or before $p$ (in the direction of $\varphi$), and $\Delta F=0$
otherwise (including when $x$ lies on another face); if $p$ and $y$ lie on different faces, then
$\Delta F=0$ when $x$ lies on one of these two faces and $\Delta F=-2$ otherwise. A move at a user is the same
computation applied to $\sigma_1^{-1}(p)$, $\sigma_1^{-1}(x)$, $\sigma_1^{-1}(y)$. Annealing accepts a move with
$\Delta F<0$ with probability $e^{\Delta F/T}$ at temperature $T$, cooled geometrically from $1.5$ to $0.05$ over
$20$ moves per dart, and is followed by first-improvement hill climbing.

\subsection{Reproducibility}
Seeds affect only tie-breaking in the shared-face search and the random corners. For each community and rule we
ran two seeds; the full range of the final $\nloc$ across seeds was at most $0.3$ under the main rule. The
shuffled ablation used two independent random orders. The bracketing samples used a fixed seed for thread
selection. The rotations of the dessins in \cref{fig:static} were checked against \cref{tab:static-rotations}, and the numbers of \cref{tab:toy-steps} are produced by the implementation.

\end{document}